\documentclass[a4paper,12pt]{article}
\PassOptionsToPackage{table}{xcolor}
\usepackage{jheppub} % for details on the use of the package, please
\usepackage[T1]{fontenc} % if needed

\usepackage{wrapfig}
\usepackage{tabu}
\usepackage{mathrsfs,amsmath,amssymb,amsthm,amsfonts,tikz,hyperref,graphicx,accents,color,bbold}
\usepackage{dsfont,epiolmec, latexsym, comment}
\usepackage{leftidx}
\usepackage{multirow}
\usepackage{tabularx}
\usepackage[utf8]{inputenc}
\usepackage{xcolor}
\usepackage{array}
\usepackage{cleveref}

\usepackage{nicematrix}

\def\be{\begin{eqnarray}}
\def\ee{\end{eqnarray}}

\definecolor{light gray}{RGB}{220,220,220}
\definecolor{dark purple}{RGB}{108,0,217}
\definecolor{pink}{RGB}{190,20,100}
\definecolor{orang}{RGB}{193,63,0}
\definecolor{green}{RGB}{11,98,17}
\definecolor{darkpink}{RGB}{153,0,76}
\definecolor{bluegreen}{RGB}{0,102,102}
\definecolor{greenlagan}{RGB}{0,102,0}
\definecolor{redgreen}{RGB}{102,102,0}
\definecolor{Redgreen}{RGB}{153,76,0}
\definecolor{vividviolet}{rgb}{0.62, 0.0, 1.0}
\definecolor{amaranth}{rgb}{0.9, 0.17, 0.31}
\definecolor{palatinateblue}{rgb}{0.15, 0.23, 0.89}
\definecolor{brightpink}{rgb}{1.0, 0.0, 0.5}
\definecolor{cornflowerblue}{rgb}{0.39, 0.58, 0.93}
\definecolor{deepcarminepink}{rgb}{0.94, 0.19, 0.22}
\definecolor{radicalred}{rgb}{1.0, 0.21, 0.37}

\usepackage[percent]{overpic}
\usepackage{wrapfig}
\usepackage{bbm}

\hypersetup{ linktoc=all,
    colorlinks, linkcolor={blue}, %{vividviolet},
    citecolor={red}, urlcolor={darkpink}
}

\graphicspath{{Images/}}

\DeclareFontFamily{OT1}{rsfs}{}
\DeclareFontShape{OT1}{rsfs}{m}{n}{ <-7> rsfs5 <7-10> rsfs7 <10->rsfs10}{} 
\DeclareMathAlphabet{\mycal}{OT1}{rsfs}{m}{n}

\title{{\LARGE{
Universal Early-Time Growth in Quantum Circuit Complexity
}}}
\author[a,b]{S. Shajidul Haque}
\author[a,c,e]{, Ghadir Jafari}
\author[d]{, Bret Underwood}

\affiliation{\it $^a$ Department of Mathematics and Applied Mathematics,
	University of Cape Town, Rondebosch, Cape Town, 7700, South Africa
}

\affiliation{\it $^{b}$ National Institute for Theoretical and Computational Sciences (NITheCS), Private Bag X1, Matieland, South Africa}

\affiliation{\it $^c$ Department of Physics Education, Farhangian University, P.O. Box 14665-889, Tehran, Iran}

\affiliation{\it $^e$ School of Astronomy, Institute for Research in Fundamental Sciences (IPM), P. O. Box 19395-5531, Tehran, Iran}

\affiliation{\it $^d$ Department of Physics, Pacific Lutheran University, Tacoma, WA 98447
}

\emailAdd{shajid.haque@uct.ac.za }
\emailAdd{ghjafari@ipm.ir}
\emailAdd{bret.underwood@plu.edu}

\abstract{
We show that quantum circuit complexity for the unitary time evolution operator of any time-independent Hamiltonian is 
bounded by linear growth at early times, independent of any choices of the fundamental gates or cost metric. Deviations from linear early-time growth arise from the commutation algebra of the gates and are manifestly negative for any circuit, decreasing the linear growth rate and leading to a bound on the growth rate of complexity of a circuit at early times. 
We illustrate this general result by applying it to qubit and harmonic oscillator systems, including the coupled and anharmonic oscillator.
By discretizing free and interacting scalar field theories on a lattice, we are also able to extract the early-time behavior and dependence on the lattice spacing of complexity of these field theories in the continuum limit, demonstrating how this approach applies to systems that have been previously difficult to study using existing techniques for quantum circuit complexity.
}    
\begin{document}

\maketitle

\section{Introduction}

Quantum circuit complexity refers to the minimum number of simple unitary operations needed to transform a reference quantum state into a target state.
A notable diagnostic within quantum information theory, quantum circuit complexity (hereafter referred to simply as complexity) has led to new insights into various fields, including quantum many-body systems \cite{Ali:2018aon, Caputa:2022yju}, quantum computation \cite{watrous2008quantum, Chapman_2022}, cosmology \cite{Bhattacharyya:2020kgu,Bhattacharyya:2020rpy, Bhattacharyya:2024duw, Haque:2021hyw}, astrophysics \cite{Dixit:2023fke}, and black hole physics \cite{Ryu:2006bv, VanRaamsdonk:2010pw, Susskind:2014moa}.
In addition, complexity can be used as a proxy for other various physical quantities, leading to its use in understanding the dynamics of quantum systems, particularly in relation to chaos \cite{Balasubramanian:2019wgd, Ali:2019zcj, Bhattacharyya:2020iic, Bhattacharyya:2020art, Bhattacharyya:2023grv, Bhattacharyya:2019txx, Kudler-Flam:2019kxq, Balasubramanian:2022tpr,Craps:2022ese,Craps:2023rur}, decoherence \cite{Haque:2021kdm, Bhattacharyya:2022rhm, Bhattacharyya:2021fii}, and phase transitions \cite{Caputa:2022yju, Beetar:2023mfn, Caputa:2022eye}.

Recently, there has also been a surge in interest in complexity stemming from its role in holography, where it is a useful probe to investigate the interior of an eternal black hole in anti-de-Sitter space (AdS).
Complexity was proposed within the AdS/CFT framework as a quantity in the conformal field theory that can help understand the growth in the size of the wormhole as the black hole thermalizes \cite{Hartman:2013qma,Maldacena:2013xja,Susskind:2014moa,Susskind:2014rva}.
Beyond the original proposals \cite{Stanford_2014,Brown:2015bva,Alishahiha:2015rta}, it has been suggested that complexity is dual to a wide range of geometric objects in the AdS bulk \cite{Belin_2022, Belin_2023, myers2024complexity}.
Because of these interesting applications, formulating and generalizing complexity for free and interacting quantum field theories can provide valuable insights into holographic and other scenarios.

While originally defined through discrete gate counting, complexity can be extended to a continuous geometric formulation \cite{Nielsen_2006, dowling2006geometry} in which complexity corresponds to the shortest path in the space of unitaries required to construct a target unitary from the identity.
Various approaches to this continuous description of complexity have been proposed (see \cite{Jefferson:2017sdb, Hackl:2018ptj, Balasubramanian:2019wgd, Ali:2018fcz}), and the sensitivity of different approaches varies, sometimes leading to qualitatively different findings \cite{Bhattacharyya:2020rpy}.
One approach characterizes the unitaries by their action on a specific Gaussian form for the wavefunction \cite{Jefferson:2017sdb, Guo:2018kzl}
which, while simplifying formulation of the problem of minimizing the path in the space of unitaries, limits the scope of this approach to specific Gaussian reference and target states.
Another approach, Krylov complexity \footnote{Recently, the relation between Krylov and circuit complexity has been studied in \cite{Craps:2023ivc}. } \cite{Rabinovici:2020ryf,Barbon:2019wsy, Parker:2018yvk,  Dymarsky:2021bjq}, is a proposal for complexity which measures the degree of operator growth in some specific basis, but is more efficiently formulated for unitaries that can be written as generalized coherent state operators \cite{Balasubramanian:2022tpr, Caputa:2021sib, Haque:2022ncl}.

In this paper we will focus instead on an operator-based approach to complexity \cite{NL1,Balasubramanian:2019wgd,Balasubramanian:2021mxo,Haque:2021hyw} which uses the Euler-Arnold equation to find the shortest path in the space of unitiaries.
This approach will be independent of the specific form of reference and target states and not limited to generalized coherent state operators, allowing us to smoothly extend our analysis to non-Gaussian evolution as well as free and interacting quantum field theories.
In this operator-complexity framework, the target unitary (which we will take to be the time-evolution operator of some generic time-independent Hamiltonian) is constructed from a set of fundamental gates that form a Lie algebra. The minimal geodesic on the associated group manifold of the Lie algebra is determined as a solution to the Euler-Arnold equation \cite{Balasubramanian:2019wgd, Balasubramanian:2021mxo,Haque:2021hyw}. 
This group-theoretic approach is advantageous because the geometry is defined by the Lie algebra generators (subject to the choice of a ``cost metric'' on the space of { operators}) and is inherently independent of the reference and target states.
Furthermore, this method can be readily be extended to study the complexity of non-Gaussian and interacting systems by considering the Lie algebra of the set of fundamental operators characterizing these systems.

%This paper is organized as follows.
After reviewing some general framework and results associated with operator complexity in Section \ref{sec:OperatorC}, we demonstrate in Section \ref{sec:pert} a perturbative approach that allows us to find generic solutions for the complexity at early times.
%focus on finding generic solutions to the Euler-Arnold equation that are perturbative in time in Section \ref{sec:pert}.
{
While some features of interest in complexity occur at late times (such as the saturation of complexity at exponentially long times for chaotic systems e.g.~\cite{Susskind:2014moa}), other features may be of particular interest at early times.
For example, some diagnostics, such as the out-of-time-order correlator, are more sensitive to the onset of scrambling or quantum chaos at early times \cite{Shenker:2013pqa,Maldacena:2015waa,Zhuang:2019jyq}, so it is useful to determine whether operator circuit complexity has a similar sensitivity.
In addition, it is interesting to explore whether expected bounds on the growth of scrambling in other diagnostics \cite{Maldacena:2013xja,Parker:2018yvk,Hornedal:2022pkc} extend to bounds on the growth of operator circuit complexity.

While we are not yet able to distinguish whether the early-time behavior of complexity we find here is sensitive to quantum chaos in similar ways, we do show in Section \ref{sec:pert} that the complexity for the time-evolution operator of a generic time-independent Hamiltonian takes a universal form which is linear at early times.
In particular, we show that the subleading correction to the early-time linear growth of complexity always appears with a minus sign, bounding the growth of complexity to be at most linear, demonstrating a general feature of complexity for any system.
This result applies independent of the algebra of the fundamental gates or the metric on the space of unitaries, allowing it to applicable to a wide range of applications in quantum mechanics and interacting quantum field theories, including integrable and chaotic models.
Note that while some previous results for operator complexity have found linear growth of complexity for certain particular systems \cite{Balasubramanian:2019wgd,Haferkamp:2021uxo}, we extend these results to find that early time linear growth is a generic feature of complexity for any quantum system with any choice of gates.
In particular, our approach allows us to explore the operator complexity of interacting systems, which have been difficult to study using other techniques.
}
We illustrate our general result by applying it to several quantum mechanical examples in Section \ref{sec:examples}, including qubits and free and coupled harmonic oscillators, as well as free and interacting scalar  field theories.
{ These examples serve to demonstrate and validate the general results of Section \ref{sec:pert}, as well as demonstrate how our approach can lead to the study of complexity in new classes of interacting models.}
The appendices include some additional details, including a derivation of the Euler-Arnold equation in Appendix \ref{app:EulerArnoldDerivation}, a generalization of our perturbative approach to time-dependent Hamiltonians in Appendix \ref{sec:TimeDepPert}, and detailed calculations for the anharmonic oscillator in Appendix \ref{app:Anharmonic}.

\section{Operator Complexity}
\label{sec:OperatorC}

We will consider a notion of continuous operator complexity following \cite{Balasubramanian:2019wgd,Balasubramanian:2021mxo,Haque:2021hyw}.
Consider a quantum circuit consisting of the transformation of a given reference state $|\psi\rangle_{\rm R}$ to a specified target state $|\psi\rangle_{\rm T}$ through some sequence of unitary operations. 
Given these reference and target states, we can represent this overall transformation with some target unitary operator $\hat{{\mathcal U}}_{\rm target}$,
\begin{equation}
    |\psi\rangle_{\rm T} = \hat {\mathcal U}_{\rm target}\ |\psi\rangle_{\rm R}\, .
\label{eq:CircuitDef}
\end{equation}
Taking the target unitary as a sequence of operations with respect to some set of fundamental Hermitian ``gate'' operators $\{\hat{\mathcal O}_I\}$, it can be written
\begin{equation}
\hat {\mathcal U}_{\rm target} = {\mathcal P}\, {\rm exp}\left[-i \int_0^1 V^I(s) \hat {\mathcal O}_I\ ds\right]\, ,
\label{eq:UtargetDef}
\end{equation}
where the { $V^I(s)$ are tangent vectors of a path in the space of operators generated by $\{\hat {\mathcal O}_I\}$,}
%specify the path of the sequence of operators, 
and the path-ordering ${\mathcal P}$ ensures that the operators are applied sequentially
%over 
{ through the circuit 
%time 
parameter\footnote{The parameter $s$ is commonly referred to as the circuit ``time,'' but we will instead refer to $s$ as the circuit parameter to avoid confusion with the physical time $t$ introduced in the target operator below. The circuit parameter $s$ has no physical meaning, since it is merely a parameter used to ensure the ordering of the operators along the path.} $s$.
%from $s = 0$ to $s = s'$.
As we will see soon, our choice for the circuit depth cost function will be $s$-reparameterization invariant, so we will order the circuit as running from $s=0$ to $s=1$ without loss of generality \cite{Nielsen:2005mkt}.}

We will take the gate operators 
to obey a Lie algebra 
$\big[\hat{\mathcal{O}}_I,\hat{\mathcal{O}}_J\big]= i f^K_{IJ} \hat {\mathcal{O}}_K$ with 
structure constants $f^K_{IJ}$. 
In this work we will take the target unitary to be the time-evolution operator associated with some time-independent Hamiltonian $\hat H$
\be
\hat {\mathcal U}_{\rm target} = e^{-it \hat H} = e^{-it\, \omega^I \hat{\mathcal O}_I}\, ,
\label{eq:UTargetTimeEvolution}
\ee
where the $\omega^I$ are the expansion of the Hamiltonian in the basis of the gate operators $\hat {\mathcal O}_I$.
More generally, the $s$-dependent unitary
\be
\hat U(s) = {\mathcal P}\, {\rm exp}\left[-i \int_0^s V^I(s') \hat {\mathcal O}_I\ ds'\right]\, ,
\label{eq:UsDef}
\ee
is a solution to the differential equation
\be
\frac{d\hat U(s)}{ds} = -i V^I(s)\ \hat {\mathcal O}_I\ \hat U(s)\, ,
\label{eq:UsDiffeq}
\ee
subject to the boundary conditions 
\be
\hat U(0) = \hat {\mathbb{1}}\, \hspace{.2in} \mbox{and} \hspace{.2in} \hat U(1) = \hat {\mathcal U}_{\rm target}\, .
\label{eq:UsBoundary}
\ee

In principle, there are many different paths (``circuits'') $V^I(s)$ that can be used to build the unitary $\hat U(s)$ subject to the boundary conditions (\ref{eq:UsBoundary}).
In order to identify the ``optimal'' path, we will characterize each path realizing the unitary (\ref{eq:UsDef}) by its circuit depth
\be
{\mathcal D}\left[V^I\right] = \int_0^1 \sqrt{G_{IJ} V^I(s) V^J(s)}\ ds\, ,
\label{eq:CircuitDepth}
\ee
corresponding to the geodesic length in the space of operators with respect to some metric $G_{IJ}$.
{ Note that the circuit depth in (\ref{eq:CircuitDepth}) is $s$-reparameterization invariant, allowing us to set $s=0$ and $s=1$ as the start and end of the circuit without loss of generality \cite{Nielsen:2005mkt}, as mentioned above.}
The metric $G_{IJ}$ identifies the operational ``cost'' or weight to building the path with any particular operator $\hat {\mathcal O}_I$.
As discussed in \cite{Haque:2021hyw}, a simple and natural choice is the flat metric $G_{IJ} = \delta_{IJ}$. Other natural choices include splitting the space into so-called ``easy'' and ``hard'' directions, the latter with a metric weighted by a cost factor \cite{NL3,Balasubramanian:2019wgd,Balasubramanian:2021mxo}. 
For now, we will assume the metric $G_{IJ}$ is constant with Euclidean signature, but otherwise unspecified.
{
In addition to the ``cost metric'' $G_{IJ}$, it is also possible to consider alternative choices for the cost functional itself inside the integral of (\ref{eq:CircuitDepth}). 
The choice in (\ref{eq:CircuitDepth}) is common and convenient, in that it yields a group manifold with Riemannian geometry. However there are other choices for $s$-reparameterization invariant cost functionals, as described in \cite{Nielsen:2005mkt}.
Previous works have found that the complexity for different cost functionals have similar features, though the scaling of the complexity with the volume may differ for field theories in the continuum limit (see \cite{Jefferson:2017sdb,Haque:2021kdm}).
We will choose our cost functional as in (\ref{eq:CircuitDepth}) for its simplicity and familiarity, though in principle our approach below should be applicable to other cost functionals as well.
}

The operator complexity for the target operator $\hat{\mathcal U}_{\rm target}$ is identified as the minimal length of the path connecting it to the unit operator
\begin{equation}
\label{compdef}
	\mathcal{C}_{\text {target }}=\min _{\left\{V^{I}\right\}} \mathcal{D}\left[V^{I}\right]=\min _{\left\{V^{I}\right\}} \int_{0}^{1} \sqrt{G_{I J} V^{I}(s) V^{J}(s)}\, d s\, ,
\end{equation}
where the minimization is taken over the possible paths $\{V^I(s)\}$ satisfying (\ref{eq:UsDiffeq}) and (\ref{eq:UsBoundary}).
The minimal path of \eqref{compdef} is a geodesic on the group manifold of the space of gates generated by $\{\hat{\mathcal O}_I\}$ with metric $G_{IJ}$ and structure constants $f^K_{IJ}$, and solves the Euler-Arnold equation\footnote{See Appendix \ref{app:EulerArnoldDerivation} for a derivation of the Euler-Arnold equation for the cost functional (\ref{eq:CircuitDepth}.}
\be
G_{IJ} \frac{dV^J}{ds} = G_{JL}f_{IK}^J V^K V^L\, .
\label{EulerArnoldEq}
\ee
In order to find the complexity for a given target operator, one must solve (\ref{EulerArnoldEq}) for the minimal paths $V^I(s)$ subject to the boundary conditions (\ref{eq:UsBoundary}) obtained from solutions to (\ref{eq:UsDiffeq}).

Minimum length geodesics of the circuit depth (\ref{compdef}) have several unique and useful properties that are independent of any particular set of gate operators or Hamiltonian. 
In particular, the magnitude of the circuit depth is constant along the minimal path which solves (\ref{EulerArnoldEq})
\begin{equation}\label{constantC}
    \frac{1}{2} \frac{d}{ds} \left(G_{IJ} V^I V^J\right) = V^I G_{IJ} \frac{dV^J}{ds} = G_{JL} f_{IK}^J V^I V^K V^L = 0\, ,
\end{equation}
which vanishes by the antisymmetry of the structure constants, so that the circuit depth is independent of the circuit parameter $s$.
Thus, the complexity is simply the value of the minimized circuit depth at any point along the circuit,
\begin{equation}
    {\mathcal C}_{\rm target} = \left.\sqrt{G_{IJ} V^I V^J}\right|_{\rm min}\, .
\end{equation}
The constancy of the circuit depth (\ref{constantC}) also implies that the acceleration $\frac{dV^I}{ds}$ of the minimal path is perpendicular to its tangent vector $V^I(s)$.

In previous studies \cite{Balasubramanian:2019wgd, Haque:2021hyw}, the quantum circuit complexity (\ref{compdef}) for special choices of target operators and Hamiltonians was found by using the explicit form of the structure constants and matrix representations for the fundamental gates $\{\hat {\mathcal O}_I\}$, and solving the resulting, relatively simple, coupled differential equations (\ref{EulerArnoldEq}) and (\ref{eq:UsDiffeq}).
In general, however, it is difficult to specify the representations for an arbitrary set of fundamental gates and structure constants $f_{IJ}^K$ and solve the resulting coupled differential equations.
In the next section, we present a perturbative approach to finding the minimal length geodesics solving 
(\ref{EulerArnoldEq}) and (\ref{eq:UsDiffeq}) subject to the boundary conditions (\ref{eq:UsBoundary}) that works for any choice of structure constants and time-independent Hamiltonian, and does not require the use of any matrix representation of the fundamental gates.

\section{A Perturbative Approximation in Physical Time}
\label{sec:pert}

As discussed in the previous section, in order to find the quantum circuit complexity for some target operator $\hat {\mathcal U}_{\rm target}$ given some fundamental gates with structure constants $[\hat {\mathcal O}_I,\hat {\mathcal O}_J] = i f_{IJ}^K \hat {\mathcal O}_K$, we need to find the minimal geodesic $V^I(s)$ satisfying
\begin{equation}\label{maineqs}
	G_{IJ}	\dot{V}^J(s)=f^J_{IK} {V}^K(s) G_{JL} {V}^L(s)\quad ; \quad  \frac{d \hat{U}(s)}{ds}=-i V^I(s) \mathcal{O}_I \hat{U}(s)\, ,
\end{equation}
subject to the boundary conditions $\hat U(0)=\hat {\mathbb{1}}$ and $\hat U(1)=\hat {\mathcal U}_{\rm target}$.
We will consider our target unitary operator to be written as a time-independent Hamiltonian time evolution operator\footnote{We extend this analysis for time-dependent Hamiltonians in Appendix \ref{sec:TimeDepPert}.} expanded over the fundamental gates 
\begin{equation}
\label{UtargetPert}
\hat{\mathcal{U}}_{\rm target}=e^{-i t \omega^I\hat {\mathcal O}_I}\, ,
\end{equation}
as in (\ref{eq:UTargetTimeEvolution}), where the $\omega^I$ are parameters that specify the Hamiltonian.
For general choices of fundamental gates and target operator, solving these coupled equations exactly will be quite difficult.
Instead, we will
expand the tangent vector $V^I(s)$ and s-dependent unitary $\hat U(s)$ perturbatively in {(physical)} time $t$ as
\be
\label{eq:Vpert}
V^I(s) &=& V_0^I(s) + t\, V_1^I(s) + t^2\, V_2^I(s) + \ldots\\
\label{eq:Upert}
\hat U(s) &=& \hat U_0(s) + t\, \hat U_1(s) + t^2\, \hat U_2(s) + \ldots 
\ee
and so on. Inserting (\ref{eq:Vpert}),(\ref{eq:Upert}) into (\ref{maineqs}), we can solve the resulting equations perturbatively in time $t$, order by order.

\subsection{Complexity at Leading and Next-to-Leading Order in Time}
\label{subsec:PertComplexity}

At zeroth order in time $t =0$, our target operator is simply the identity operator $\hat {\mathcal U}_{\rm target} = \hat{\mathbb{1}}$. The only solution to (\ref{maineqs}) subject to the boundary conditions, then, is the trivial one $V^I_0(s) = 0$ and $\hat U_0(s) = \hat{\mathbb{1}}$.
At first order in time $t$, the equations (\ref{maineqs}) become
\begin{equation}\label{dteqs}
     \dot{V}_1^I(s)=0  \quad; \quad \frac{d \hat{U}_1(s)}{ds}=-i V_1^I(s) \mathcal{O}_I \, .
     \end{equation}
The solutions, subject to $\hat U(0) = \hat {\mathbb{1}}$, are
    \be
        \label{V1}
        V_1^I(s) &=& \lambda_1^I\, ; \\
        \label{U1}
        \hat U_1(s) &=& {-i s \lambda_1^I \hat {\mathcal O}_I}\, ,
    \ee
for some integration constants $\lambda_1^I$. 
By matching the solution (\ref{U1}) to the boundary condition $\hat U(1) = \hat {\mathcal U}_{\rm target} \approx \hat {\mathbb{1}} - it\, \omega^I\, \hat {\mathcal O}_I$ expanded to first order, the integration constants are determined to be the expansion parameters of the Hamiltonian $\lambda^I = \omega^I$.
The resulting circuit complexity, at leading order in time $t$, is then
\be
\label{C1}
{\mathcal C} = \sqrt{G_{IJ} V^I(s) V^J(s)} \approx t \sqrt{G_{IJ} V_1^I(s) V_1^J(s)} = |\omega|t\, ,
\ee
where $|\omega|=\sqrt{G_{IJ}\omega^I\omega^J}$ is the magnitude of the Hamiltonian expanded in the gate basis. This result, valid at leading order in time, demonstrates \emph{universal early-time linear growth of quantum circuit complexity for \underline{any} fundamental gate set}.
Using the expansion $\hat H = \omega^I \hat{\mathcal O}_I$ and $\operatorname{tr}(\hat{\mathcal{O}}_I \hat{\mathcal{O}}_J)=G_{IJ}$, we can write $\omega^I=G^{IJ}\operatorname{tr}( \hat{\mathcal{O}}^{\dagger}_J \hat H)$, 
so that the universal early-time growth of complexity takes the compact form
\begin{equation}
    {\mathcal C} \approx |\operatorname{tr}(\hat{\mathcal{O}}^{\dagger}_I \hat H)|\, t\, .
\end{equation}

At second order in time $t$ the equations (\ref{maineqs}) become
\begin{align}
\label{pereqs2}
& G_{IJ}\dot{V}_2^J(s)=f^J_{IK}\omega^K G_{JL} \omega^{L}\, ;\\  
\label{U2eqn}
& \frac{d \hat{U}_2(s)}{ds}=-i  V_2^I(s) \mathcal{O}_I-s (\omega^I \mathcal{O}_I )^2\, ,
\end{align}
where we used the first-order solutions (\ref{V1}) and (\ref{U1}).
The solutions to \eqref{pereqs2} and \eqref{U2eqn} are
\be
\label{solv2}
V_2^I(s) & =& s \Omega_2^I+\lambda_2^I\, ; \\
\label{solu2}
\hat{U}_2(s) &=& -\tfrac12 s^2 ( i  \Omega_2^I \hat{\mathcal{O}}_I+(\omega^I \hat{\mathcal{O}}_I )^2)-i s \lambda_2^I\hat{\mathcal{O}}_I\, .
\ee
where the $\lambda_2^I$ are some integration constants,
and $\Omega_2^I$ is a constant vector
\begin{equation}
\label{Omega2}
    \Omega_2^I=G^{LI}f^J_{LK}\omega^K G_{JM} \omega^{M}\, ,
\end{equation}
which depends on the structure constants and the Hamiltonian (through the $\omega^I$).
In order to determine the constants $\lambda_2^I$ of integration, we will match the solution (\ref{solu2}) at $s=1$ to the target operator $\hat {\mathcal U}_{\rm target}$ (\ref{eq:UTargetTimeEvolution}) expanded to second-order in $t$ 
\begin{equation}
    \hat {\mathcal U}_{\rm target,2} = - \tfrac12{ t}^2 (\omega^I\hat {\mathcal O}_I)^2\, .
\end{equation}
The boundary condition requires that the integration constants take the value
\begin{equation}
\label{kappa}
    \lambda_2^I=-\tfrac12 \Omega_2^I \, ,
\end{equation}
so that the solutions (\ref{solv2}),(\ref{solu2}) for the geodesic and unitary at second order in time become
\begin{align}
\label{v2solution}
     &V_2^I(s)=(s-\tfrac12)\Omega_2^I\, ;\\
\label{U2ssolution}
     & \hat{U}_2(s)=-i\tfrac12 s(s-1) \Omega_2^I\mathcal{O}_I-\tfrac{s^2}{2}(\omega^I\mathcal{O}_I)^2\, .
\end{align}

While it is tempting to use (\ref{v2solution}) to evaluate the complexity at next-to-leading order in time $t$, this is not sufficient. Including contributions to the tangent vector up to second order, we have (we will consider the \emph{square} of complexity here, for notational simplicity)
\be
{\mathcal C}^2 &=& G_{IJ} V^I V^J \approx G_{IJ} \left(t\, V_1^I + t^2\, V_2^I\right)\left(t\, V_1^J + t^2\, V_2^J\right) \nonumber \\ 
\label{C2false}
    &\approx & t^2\, G_{IJ} V_1^I V_1^J + 2 t^3 G_{IJ} V_1^I V_2^J + t^4 G_{IJ} V_2^I V_2^J\, .
\ee
The first term in (\ref{C2false}) gives the leading order result for the complexity (\ref{C1}) when evaluated on the solution (\ref{V1}). 
The next-to-leading order term contributes at power ${\mathcal O}(t^3)$ but vanishes
\begin{equation}
    G_{IJ} V_1^I V_2^J = G_{IJ} \omega_2^I \Omega^J = G_{IJ} \omega^I G^{LJ} f^N_{LK}\omega^K G_{NM} \omega^{M} = f^N_{LK}\omega^L \omega^{K} G_{NM}  \omega^N =0\, ,
\end{equation}
because of the antisymmetry of the structure constants $f_{LK}^N$.
Thus, the only non-vanishing contribution to (\ref{C2false}) occurs at order ${\mathcal O}(t^4)$.
However, in order to correctly capture the contribution (\ref{C2false}) at order ${\mathcal O}(t^4)$, we must include contributions from the \emph{next} order in perturbation theory\footnote{It is also possible to see that the last term of (\ref{C2false}) evaluated on the second-order solution (\ref{v2solution}) will be explicitly $s$-dependent, in contradiction to the general result (\ref{constantC}) found in the previous section that the minimized circuit depth must be $s$-independent. We must go to next order in perturbation theory in order to resolve this.}, 
$V^I(s) \approx t\, V_1^I(s) + t^2\, V_2^I(s) + t^3\, V_3^I(s)$, so that (\ref{C2false}) receives another contribution at ${\mathcal O}(t^4)$
\be
\label{C3}
{\mathcal C}^2 &\approx & t^2\, G_{IJ} V_1^I V_1^J + t^4 G_{IJ} V_2^I V_2^J + 2 t^4\, G_{IJ} V_1^I V_3^J\, ,
\ee
and we dropped higher order terms.

Considering the tangent vector and unitary at third-order, then
\be
V^I(s) &\approx & t\, V_1^I(s) + t^2\, V_2^I(s) + t^3\, V_3^I(s)\, ' \\
\hat U(s) &\approx & \hat U_0(s) + t\, \hat U_1(s) + t^2\, \hat U_2(s) + t^3\, \hat U_3(s)\, ,
\ee
the equations (\ref{maineqs}) become
\be
\label{V3eqn}
G_{IJ}\frac{d}{ds}V_3^J(s) &=& f^J_{IK} V_2^K(s) G_{JL} V_1^L(s)+f^J_{IK} V_1^K(s) G_{JL} V_2^L(s)\, ; \\
\label{U3eqn}
\frac{d}{ds}\hat{U}_3(s) &=& -i  V^I_3(s) \hat {\mathcal O}_I -i  V^I_2(s) \hat {\mathcal O}_I \hat{U}_1(s)-i  V^I_1(s) \hat {\mathcal O}_I \hat{U}_2(s)\, .
\ee
Using the solutions for $V_1^I(s)$ (\ref{V1}) and $V_2^I(s)$ (\ref{v2solution}), (\ref{V3eqn}) has the solution
\begin{equation}
V_3^I = s(s-1)\Omega_3^I+\lambda_3^I\, ,
\end{equation}
where $\lambda_3^I$ is another integration constant and
\be
\Omega_3^I= G^{MI}f^J_{MK} {\Omega_2}^{(K} G_{JL} {\omega}^{L)}\, ,
\ee
with parentheses denoting symmetrization of the indices.
The solution to (\ref{U3eqn}) for the unitary, using (\ref{U1}) and (\ref{U2ssolution}), becomes
\begin{align}
\label{U3}
    \hat{U}_3(s) = &-is\lambda_3^I \hat {\mathcal O}_I- i \left(\frac{s^3}{3}-\frac{s^2}{2}\right)\Omega_3^I \hat {\mathcal O}_I -\left(\frac{s^3}{3}-\frac{s^2}{4}\right)\omega^J \Omega_2^I \hat {\mathcal O}_I \hat {\mathcal O}_J\\
            &-\tfrac12 \left(\frac{s^3}{3}-\frac{s^2}{2}\right)\omega^J \Omega_2^I \hat {\mathcal O}_J \hat {\mathcal O}_I + i\tfrac{s^3}{6}(\omega^I\hat {\mathcal O}_I)^3\, .
\end{align}
In order to find the value of the integration constant $\lambda_3^I$, we must evaluate (\ref{U3}) at $s=1$ 
\begin{align}
 \hat{U}_3(1)= &-i\lambda_3^I\hat {\mathcal O}_I+  \tfrac{i}{6} \Omega_3^I \hat {\mathcal O}_I-\tfrac{1}{12} \omega^J \Omega_2^I[\hat {\mathcal O}_I,\hat {\mathcal O}_J] +\tfrac{i}{6}(\omega^I\hat {\mathcal O}_I)^3\nonumber\\
        &=-i\lambda_3^I\hat {\mathcal O}_I+  \tfrac{i}{6} \Omega_3^I \hat {\mathcal O}_I-\tfrac{i}{12} \omega^J \Omega_2^I f^K_{IJ}\hat {\mathcal O}_K+\tfrac{i}{6}(\omega^I\hat {\mathcal O}_I)^3\, ,
\end{align}
and match it to the target unitary (\ref{UtargetPert}) expanded to third order in time
\begin{equation}
    \hat {\mathcal U}_{\rm target, 3} = \tfrac{i}{6}(\omega^I\hat {\mathcal O}_I)^3\, .
\end{equation}
The integration constant is found to be
 \begin{equation}
     \lambda_3^I =\tfrac{1}{6}\Omega_3^I-\tfrac{1}{12} \omega^J \Omega_2^K f^I_{KJ}\, ,
 \end{equation}
so that finally the third-order contribution to the tangent vector is
\begin{equation}
\label{V3solution}
    V_3^I= \tfrac1{6} (6s(s-1)+1)\Omega_3^I  -\tfrac{1}{12} \omega^J \Omega_2^K f^I_{KJ}\, .
\end{equation}

Evaluating (\ref{C3})
\be
\label{C4}
{\mathcal C}^2 &\approx & t^2\, G_{IJ} V_1^I V_1^J + t^4 G_{IJ} V_2^I V_2^J + 2 t^4\, G_{IJ} V_1^I V_3^J\, ,
\ee
with the solutions for the tangent vectors order-by-order (\ref{V1}), (\ref{v2solution}), (\ref{V3solution}), and making use of the relations
\begin{align}
 \omega^J \Omega_2^K f^I_{KJ} G_{IL}\omega^L&=G_{IJ} {\Omega_2}^{I}{\Omega_2}^{J} = |\Omega_2|^2\, ; \nonumber \\
 G_{IJ}   \omega^J \Omega_3^I&= \tfrac12\omega^{I}f^J_{IK} {\Omega_2}^{K} G_{JL} {\omega}^{L}=-\tfrac12 |\Omega_2|^2\, ,
\end{align}
we find that the contributions to (\ref{C4}) at ${\mathcal O}(t^4)$ become
\be
G_{IJ}V_2^IV_2^J + 2G_{IJ}V_1^IV_3^J=\left[(s-\frac{1}{2})^2-s (s-1)-\tfrac13\right]|\Omega_2|^2=-\tfrac{1}{12}|\Omega_2|^2\, ,
\ee
so that the (squared) complexity is
\begin{equation} 
    \mathcal{C}^2=t^2\, G_{IJ}\omega^I\omega^J -t^4\, \tfrac{1}{12}|\Omega_2|^2 + \ldots
\end{equation} 
Note that the result is $s$-independent after including the third-order solution for the tangent vector $V_3^I$, consistent with the result (\ref{constantC}) that the circuit depth evaluated on the minimized geodesic must be $s$-independent.
Taking the square root, the complexity for any unitary evaluated up to next-to-leading order in time takes the universal form
\begin{equation}
    \label{C22}
	\mathcal{C}\approx  t |\omega|\left(1-\tfrac1{24} t^2\frac{|\Omega_2|^2}{|\omega|^2}\right)\, .
\end{equation}
This is the main result of this paper: the complexity of {\it any} quantum circuit realizing the target unitary (\ref{UtargetPert}) with the geodesic length depth functional (\ref{eq:CircuitDepth}) is of the universal form (\ref{C22}) at leading and next-to-leading order in time.

\begin{figure}[t]
\centering\includegraphics[width=0.7\textwidth]{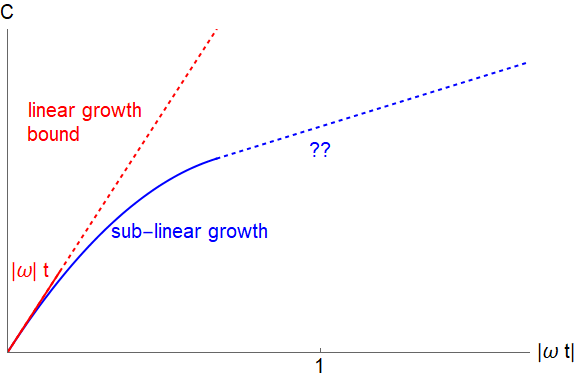}
\caption{The generic behavior of the operator circuit complexity of the time evolution operator at early times $|\omega|t\ll 1$ for a generic cost metric and set of gates, (\ref{C22}). The complexity is bounded from above by the linear growth of complexity at early times, ${\mathcal C}\approx t\, |\omega|$ (red lines). Non-trivial commutation relations between the gates leads to subleading behavior for the complexity that reduces the growth of complexity (blue solid line). At late times, the perturbative approach fails to capture the behavior of complexity (dashed blue lines).}
\label{fig:ComplexityGraph}
\end{figure}

As noted earlier, the leading-order behavior of the complexity (\ref{C22}) for any unitary is linear in time and is independent of the group structure of the fundamental gates $\{\hat{\mathcal O}_I\}$, insofar as a different set of fundamental gates will lead to a different decomposition $\omega^I$ of the Hamiltonian.
Because the vectors $\omega^I, \Omega_2^I$ are real vectors, the sign of the subleading-order term in (\ref{C22}) is fixed and thus the complexity is \emph{bounded by linear growth} at early times, so that any corrections to the leading-order behavior reduce the growth rate of complexity. This is again true for any set of fundamental gates or target unitary operator.
{ We note that while \cite{Balasubramanian:2019wgd} found linear growth in time for the special cases of $\mathfrak{su}(2)$ and $\mathfrak{su}(N)$, our result finds that linear growth is a generic feature of the complexity of the time-evolution operator at early times for any system of gates and any cost metric. Moreover, the sign of the subleading term in \ref{C22} implies that this linear growth bounds the early-time complexity from above.
The generic behavior of the complexity at early times from (\ref{C22}) is shown in Figure \ref{fig:ComplexityGraph}.}

The subleading term in (\ref{C22}) depends on the algebra of the gates through the structure constants $f_{IJ}^K$ in $\Omega_2^I$, and we can see the conceptual origin of this term as arising from the algebra of the gates generating new terms through $[\hat {\mathcal O}_I, \hat {\mathcal O}_J] = i f_{IJ}^K \hat {\mathcal O_K}$.
As a simple example, consider a simple circuit consisting of two straight-line segments 
\be
V^I(s) = \begin{cases}
    V_1^I & 0 \leq s < s_1 \cr
    V_2^I & s_1 \leq s \leq 1 
\end{cases}
\ee
where $V_1^I, V_2^I$ are $s$-independent vector components.
The corresponding unitary constructed from this circuit would be
\begin{align}
{\mathcal P}\, {\rm exp}\left[-i \int_0^1 V^I(s) \hat {\mathcal O}_I\ ds\right] &= e^{-i (1-s_1) V_2^I \hat {\mathcal O}_I } e^{-i s_1\, V_1^I \hat {\mathcal O}_I} \nonumber \\
& = e^{-i\left((1-s_1) V_2^I + s_1 V_1^I\right)\hat {\mathcal O}_I - \frac{i}{2}s_1(1-s_1) V_2^I V_1^J f_{IJ}^K \hat {\mathcal O}_K + \ldots}
\end{align}
where the $\ldots$ represents terms with higher nested commutators of the gates.
We see that the algebra of the gates adds additional terms to the construction of the circuit, reducing the overall number of gates needed to construct the unitary.
While this simple example with piece-wise straight line segments does not minimize the circuit depth in general, it shows how the algebra of the gates can generate additional terms in the construction of the circuit, thus reducing the overall length of the circuit needed to construct the target unitary and explaining the sign of the sub-leading term in (\ref{C22}).

The leading-order and subleading-order behavior is not, however, independent of the cost $G_{IJ}$ of using particular operators (although the linear growth and sign of the correction term (\ref{C22}) are still fixed).
As an example, consider a diagonal cost metric in which some gates are ``easy'' with unit cost $G_{ij} = \delta_{ij}$, while other directions are ``hard'' $G_{\alpha\beta} = \mu\, \delta_{\alpha \beta}$ with a penalty factor $\mu \gg 1$. Separating the target operator parameters into easy and hard directions $\omega^I = \{\omega^i,\omega^\alpha\}$, we see from the solution (\ref{V1}) that while the optimal path at early times is the straight-line path
\be
\label{Veasy}
V^i(s) &\approx & V_1^i(s)\, t = \omega^i\, t\,; \\
V^\alpha(s) &\approx & V_1^\alpha(s)\, t = \omega^\alpha\, t\, ,
\ee
the complexity is dominated by the hard directions
\be
{\mathcal C} \approx t \sqrt{G_{IJ}\omega^I \omega^J} \approx \sqrt{\mu}\, t\, |\omega_{\rm hard}|\, ,
\ee
where $|\omega_{\rm hard}|^2 = \sum_\alpha \left(\omega^\alpha\right)^2$ is the magnitude of the target operator parameters in the hard directions.
It may seem surprising that the complexity is proportional to the cost of the hard directions, because as the cost becomes large the optimal geodesic 
ought to avoid the hard directions directly.
Nevertheless, we see that for very small distances in operator space the straight-line solution is the optimal one.
At next-to-leading order (that is, for later times), however, the optimal path (\ref{v2solution})
\be
V^I(s) \approx V_1^I(s)\, t + V_2^I(s)\, t^2 = \omega^I\, t + \left(s-\frac{1}{2}\right) \Omega_2^I\, t^2\, ,
\ee
becomes curved
\be
V^i(s) & \approx & \omega^i\, t+ \mu\, t^2 \left(s-\frac{1}{2}\right)\delta^{\ell i}\delta_{\kappa\lambda} \left(f_{\ell j}^\kappa \omega^j + f_{\ell\delta}^\kappa \omega^\delta\right) \omega^\lambda \equiv \omega^i\, t + \mu\, t^2 \left(s-\frac{1}{2}\right) \hat \Omega_2^i\,; 
\label{veasy}\\
V^\alpha(s) & \approx & \omega^\alpha\, t + t^2 \left(s-\frac{1}{2}\right)\delta^{\alpha\lambda}\delta_{\kappa\delta} \left(f_{\lambda j}^\kappa \omega^j + f_{\lambda \nu}^\kappa  \omega^\nu\right)\omega^\delta\, \equiv \omega^\alpha\, t + t^2 \left(s-\frac{1}{2}\right) \hat \Omega_2^\alpha\, , \label{vhard}
\ee
where we substituted in expressions for $\Omega_2^I$ and $\hat \Omega_2$ has any factors of $\mu$ scaled out.
For a large penalty factor $\mu\gg 1$, the second term (\ref{veasy}) causes the optimal path to become preferentially curved in the {\it easy} directions compared to the hard directions.
This reflects the synthesis of the hard gates indirectly through the algebra of the easy gates $\left[\hat {\mathcal O}_{\rm easy},\hat {\mathcal O}_{\rm easy}\right] \sim i f_{IJ}^K \hat {\mathcal O}_{\rm hard}$.
The resulting complexity (\ref{C22}) deviates significantly from linear growth at later times
\be
{\mathcal C} \approx \sqrt{\mu}\, t |\omega_{\rm hard}| \left[1 - \frac{1}{24}\mu \frac{\delta_{ij} \hat \Omega_2^i \hat \Omega_2^j}{|\omega_{\rm hard}|^2}t^2\right]\, ,
\label{Chard}
\ee
beyond which we lose perturbative control of the solutions.
These results are in general agreement with the conceptual reasoning of \cite{Susskind_Universality} for generic circuits.
{ In particular, \cite{Susskind_Universality} conjectures that the early time complexity should demonstrate linear growth dominated by the penalty direction, followed by a period of sub-linear growth, in excellent agreement with the generic result (\ref{Chard}).}

Finally, we expect the perturbative result (\ref{C22}) to be a good approximation up to times
\be
t_{\rm pert}\sim \frac{|\omega|}{|\Omega_2|} \sim |\omega|^{-1}\, ,
\label{eq:tpert}
\ee
since $\Omega_2^I$ (\ref{Omega2}) is quadratic in the $\omega^I$, and we assumed the structure constants $f_{IJ}^K\sim {\mathcal O}(1)$.
If the group manifold associated with the gates $\{ \hat{\mathcal O}_I\}$ has conjugate points along some direction 
associated with a particular component $\omega^{\hat I}$ of the Hamiltonian, we expect the target unitary to reach those conjugate points at the timescale
$t_{\rm conj}\sim (\omega^{\hat I})^{-1}$.
When the decomposition of the Hamiltonian into the $\omega^I$ has only a single component (or is dominated by the conjugate point direction), these two timescales are roughly equivalent, so that we expect (\ref{C22}) to be a good approximation up to the conjugate point timescale, leading to a generic expectation of linear growth up until an encounter with a conjugate point.
More generally, we expect the timescale (\ref{eq:tpert}) for the perturbative result (\ref{C22}) to be a good approximation to be less than, or the same order as, the conjugate point timescale, $t_{\rm pert} \lesssim t_{\rm conj}$.

\subsection{Perturbative Solutions at Any Order}
\label{sec:PertSolnsN}

Following the procedure of the previous subsection, it is possible to calculate the circuit complexity of a target unitary to {\bf any order in perturbation theory} in operator time $t$.
The target unitary is expanded as
\be
\hat {\mathcal U}_{\rm target} = e^{-it \omega^I \hat {\mathcal O}_I} = \sum_{n=0} \frac{(-i)^n\, t^n}{n!} \left(\omega^I \hat {\mathcal O}_I\right)^n\, ,
\ee
while the tangent vector and unitary can be similarly expanded as
\begin{equation}
	V^I(s)=\sum_{n=0} t^n V_n^I(s)\quad;\quad U=\sum_{n=0} t^n U_n(s)\, .
\end{equation}
The equations (\ref{maineqs}) then become, order-by-order,
\begin{align}
    &G_{IJ}	\frac{d}{ds}V_n^J(s)=\sum_{p+q=n} f^J_{IK} V_p^K(s) G_{JL} V_q^L(s)\, ;\nonumber \\
\label{pereqsUV}
    &\frac{d}{ds}\hat{U}_n(s)=-i \sum_{p+q=n} V^I_p(s) \hat {\mathcal O}_I \hat{U}_q(s)\, .
\end{align}
The solutions can be written as a power series in circuit parameter $s$
\begin{equation}
    \label{pertPowerSeries}
	 V_n^I(s)=\sum_{\alpha=0}^n s^{\alpha}  V_{n,\alpha}^I\quad; \quad 	 \hat{U}_n(s)=\sum_{\alpha=0}^n s^{\alpha}  \hat U_{n,\alpha}\, ,
\end{equation}
where the $V_{n,\alpha},\hat U_{n,\alpha}$ are $s$-independent constants and operators, respectively.
Inserting the power series expansion (\ref{pertPowerSeries}) into (\ref{pereqsUV}) gives a set of recursion relations for the power series coefficients of the tangent vector
\begin{align}
\label{pertRecursiveV}
    G_{IJ} V_{n,\alpha}^J=\frac{1}{\alpha}\sum_{p+q=n}\sum_{\mu+\nu=\alpha-1} f^J_{IK} {V_{p,\mu}}^K G_{JL} {V_{q,\nu}}^{L}\, ,
\end{align}
and the unitary
\begin{equation}
\label{pertRecursiveU}
    \hat U_{n,\alpha}=-i\frac{1}{\alpha}\sum_{p+q=n}\sum_{\mu+\nu=\alpha-1} {V_{p,\mu}}^I \hat {\mathcal O}_I {\hat U_{q,\nu}}\, .
\end{equation}
The boundary conditions at $s=0$ allow us to start the recursion relations with $V_{0,0}^I = 0$ and $\hat U_{0,0} = \hat {\mathbb{1}}$, while the boundary conditions at $s=1$, namely $\hat U(1) = \hat {\mathcal U}_{\rm target}$, fix the constants $V_{s,\alpha}^I$ in terms of the Hamiltonian expansion parameters $\omega^I$ through
\begin{equation}
\label{pertBoundary}
	\frac{{(-i)}^n}{n!} (\omega^I\hat {\mathcal O}_I)^n=\hat{U}_n(1)=\sum_{\alpha}^n \hat U_{n,\alpha}\, .
\end{equation}
In principle, the recursion relations (\ref{pertRecursiveV}), (\ref{pertRecursiveU}) and (\ref{pertBoundary}) can be solved for any given set of Hamiltonian expansion parameters $\omega^I$ and structure constants $f_{IJ}^K$ up to any order in time $t$.

\section{Examples}
\label{sec:examples}

In the previous section we outlined a technique for finding the optimal quantum circuit (and corresponding complexity) that constructs a target operator given a fundamental gate set by solving the Euler-Arnold equations (\ref{EulerArnoldEq}) subject to the boundary conditions (\ref{eq:UsBoundary}) perturbatively in time, to any order.
We presented explicit solutions at leading and next-to-leading order in time, finding a universal form for the complexity (\ref{C22}) valid perturbatively in time.

In this section, we examine these perturbative results for specific gate sets and target operators. %in the context of known exact solutions.
{ In cases where the exact solution is known, we will be able to validate that our perturbative approach gives the correct early-time behavior.
In cases where it is otherwise difficult to calculate the exact solution, we will show how our approach allows us to calculate the early-time behavior of complexity, and demonstrate in specific cases the general, qualitative features discussed in Section \ref{sec:pert}.
We will also demonstrate how our analysis allows the study of complexity in interacting models, such as coupled and self-interacting oscillators, which allows us to extract the early-time complexity of lattice models of free and interacting scalar fields.}
%We also apply the perturbative results to calculate the leading-order behavior of complexity for interacting quantum oscillators.

\subsection{Qubits and SU(N)}

We will begin by considering quantum circuits generated by unitary operators acting on single qubits, so that target operators $\hat {\mathcal U}_{\rm target}$ belong to the group U(2). Restricting to SU(2), the corresponding gates $\{\hat {\mathcal O}_I\} = \{\hat e_I\}$ belong to the Lie group $\mathfrak{su}(2)$ with algebra,
\be
 [\hat e_1, \hat e_2] = i \hat e_3,\ [\hat e_3, \hat e_1] =  i \hat e_2, \ 
 [\hat e_2, \hat e_3] =  i\hat e_1 \,.
\label{eq:SU2LieAlgebra}
\ee
More generally, target operators acting on $N$-qubits belong to the group ${\rm SU}\left(2^N\right)$ and are generated by the Lie algebra $\mathfrak{su}\left(2^N\right)$ with totally antisymmetric structure constants
\be
[\hat e_i,\hat e_j] = i \epsilon_{ijk}\, \hat e_k\, .
\label{eq:SUNLieAlgebra}
\ee

For an operator-space metric proportional to the identity metric $G_{IJ}\sim \delta_{IJ}$, the Euler-Arnold equation (\ref{EulerArnoldEq}) for the optimal path in ${\rm SU}\left(2^N\right)$ is solved exactly by the straight-line solution $V^I(s) = v^I$ because of the antisymmetry of the structure constants. 
The boundary condition for a target operator of the form (\ref{UtargetPert}) leads to the identification for the circuit parameters $v^I = t\omega^I $, and a corresponding complexity that is linear in time 
\be
{\mathcal C}_{\rm su(2^N)}^{(\rm exact)} = t|\omega| \, ,
\ee
up to the presence of conjugate points (see \cite{Balasubramanian:2019wgd,Balasubramanian:2021mxo,Basteiro:2021ene} for more discussion of conjugate points in ${\rm SU}\left(2^N\right)$).
This matches exactly the perturbative result (\ref{C22})
\be
{\mathcal C}^{\rm (pert)}_{\rm su(2^N)} \approx t|\omega|\left(1-\frac{1}{24}\frac{|\Omega_2|^2}{|\omega|^2}\right) = t|\omega|\, ,
\ee
since the $\Omega_2^I$ vanish identically in this case. 
Thus, the perturbative result in this case matches the exact result.

For a generic metric $G_{IJ}$, however, the solution to the Euler-Arnold equation is non-trivial and may be difficult to find analytically. Let us consider instead a simpler case, following the discussion in the previous section, in which some directions (say, ''$k$-local'' gates consisting of $n \leq k$ products of single qubit gates) have unit cost $G_{ij} = \delta_{ij}$ for $i,j \leq k$, while other directions labeled by $\alpha,\beta \geq k$ have a large cost associated to them $G_{\alpha\beta} = \mu \delta_{\alpha \beta}$, with $\mu \gg 1$, so that the metric takes block-diagonal form
\be
G_{IJ} = \begin{pNiceMatrix}
1 &  & & & & \\
    & \Ddots & & & & \\
    & & 1 & & & \\
    & & & \hspace{.1in}\mu & & \\
    & & & & \Ddots & \\
    & & & & & \mu
\end{pNiceMatrix}\, .
\ee
Even with this weighted metric, the antisymmetry of the structure constants (\ref{eq:SUNLieAlgebra}) simplifies the form of the perturbative result (\ref{C22}) to take the form
\be
{\mathcal C} \approx \sqrt{\mu}\, t\ |\omega_{\rm hard}| \left[1-\frac{1}{24} \frac{|\epsilon_{ij\alpha} \omega^j \omega^\alpha|}{|\omega_{\rm hard}|^2}\mu\ t^2\right]\, ,
\ee
where $|\omega_{\rm hard}|^2 = \sum_\alpha (\omega^\alpha)^2$ is the magnitude of the target operator in the hard directions, as before.
We see similar behavior as the earlier result (\ref{Chard}), which was derived for any set of gates: the initial behavior of complexity for any circuit built out of $\mathfrak{su}(N)$ gates initially grows linearly with time, at a rate proportional to the cost factor $\sqrt{\mu}$ and the magnitude of the target operator parameters $|\omega_{\rm hard}|$. Soon, however, it becomes easier to build the components of the target operator that are ``hard'' by using the algebra of the ``easier'' gates. For example, a simple circuit consisting of the successive application of two ``easy'' gates generates a ``hard'' gate through the algebra,
\be
e^{i \alpha\hat e_i} e^{i\beta\hat e_j} = e^{i (\alpha \hat e_i + \beta \hat e_j) -\frac{i}{2} \alpha \beta\epsilon_{ij\alpha} \hat e_\alpha + ...}
\label{eq:EasyHardSU_N}
\ee
for some specific $i,j$ in the easy directions and some $\alpha$ in the hard directions. 
Because of this effect, the rapid linear growth soon shuts off at a timescale of order $t_{\rm pert} \sim (\sqrt{\mu} |\omega_{\rm hard}|)^{-1}$, which is parametrically small for $\mu \gg 1$.
However, the complexity for $\mathfrak{su}(N)$ also has a new behavior: if the target operator is only ``hard'', so that  the $\{\omega^i\}$ vanish, then complexity reduces to pure linear growth without corrections. This seems consistent with the qualitative behavior discussed above: if, for example, there are no easy components to the target operator, then it will not be possible to use the easy directions in the quantum circuit to build the hard components of a target operator through their algebra as in (\ref{eq:EasyHardSU_N}).

In the case of $\mathfrak{su}(2)$, we can be even more explicit. Assigning the gate $\hat e_3$ to be the ``hard'' direction, the complexity at early times becomes
\be
{\mathcal C}_{\rm su(2)} \approx \sqrt{\mu}\ \omega^3\ t \left[1-\frac{1}{24} \left((\omega^1)^2 + (\omega^2)^2\right)\mu t^2\right]\, ,
\ee
so that the correction to the rapid linear growth becomes independent of $\omega^3$, the target operator in the hard direction.

\subsection{Harmonic Oscillator}
\label{sec:HarmonicOscillator}

The bosonic analog of unitary operators built using $\mathfrak{su}(2)$ gates are the creation and annihilation operators of the quantum harmonic oscillator $\hat a, \hat a^\dagger$ with $[\hat a,\hat a^\dagger] =1$
\begin{equation}
	 \hat e_1 = \frac{\hat a^2 + \hat a^{\dagger 2}}{4}, \hspace{.2in} \hat e_2 = \frac{i\left(\hat a^2 - \hat a^{\dagger 2}\right)}{4}, \hspace{.2in} \hat e_3 = \frac{\hat a \hat a^\dagger + \hat a^\dagger \hat a}{4}\, .
\end{equation}
These operators satisfy the $\mathfrak{su}(1,1)$ Lie algebra,
\begin{equation}
	 \label{su11algebra}
 [\hat e_1, \hat e_2] = - i \hat e_3,\ [\hat e_3, \hat e_1] =  i \hat e_2, \ 
 [\hat e_2, \hat e_3] =  i\hat e_1 \,,
\end{equation}
 with structure constants:
\begin{equation}
	 f^3_{12}= -1, \  f^2_{31}= 1, f^1_{23}= 1\, .
 \label{eq:su11Fij}
\end{equation}
Target operators built from these gates belong to the group\footnote{Alternatively, the gates can be rewritten as generators of $SP(2,\mathbb{R})$.} $SU(1,1)$
\be
\hat {\cal U}_{\rm target} = e^{-i \omega^1 t\hat e_1 - i \omega^2 t\hat e_2 - i \omega^3 t\hat e_3 }\, ,
\label{eq:SU11Target}
\ee
which can also be parameterized in terms of the squeezing and rotation operators
\be
\hat {\cal U}_{\rm target} = \hat S(r,\phi) \hat R(\theta)\, ,
\label{eq:SU11TargetSqueeze}
\ee
with
\be
\label{eq:SqueezeDefS}
\hat S(r, \phi) &=& e^{\frac{r}{2} (e^{- 2 i \phi} \hat a^2 - e^{2 i \phi} \hat a^{\dagger 2})} = e^{-2ir(\sin(2\phi)\hat e_1 + \cos(2\phi) \hat e_2)}\, ; \\
\hat R(\theta) &=& e^{- i \theta  \frac{\hat a^{\dagger} \hat a+\hat a \hat a^\dagger}{2}} = e^{-2i\theta \hat e_3}\, ,
\label{eq:SqueezeDefR}
\ee
where $r = r(t)$, the squeezing parameter, characterizes the amount of squeezing, $\phi = \phi(t)$ is the squeezing angle, and $\theta = \theta(t)$ is the rotation angle.

Solutions for the complexity for this set of gates for a metric proportional to the identity matrix are already known \cite{Haque:2021hyw} (see also \cite{Chowdhury:2023iwg}), so we will use this setup as another test-case to compare the perturbative result (\ref{C22}) to the known solutions. 
We can then explore the behavior of solutions with non-trivial metrics and weightings, beyond those considered in \cite{Haque:2021hyw, Chowdhury:2023iwg}.
As a simple check to start with, let us consider an equal-weight metric $G_{IJ} \sim \delta_{IJ}$ and a target operator along the direction $\omega^I = (0,0,\alpha)$ 
\be
\hat {\mathcal U}_{\rm target} = e^{-i \alpha\, t\, \hat e_3 }\, ,
\ee
corresponding to the Hamiltonian time-evolution for a free harmonic oscillator (or alternatively in the language of (\ref{eq:SU11TargetSqueeze}), zero squeezing $r=0$ and a linearly growing rotation angle $\theta = \tilde\omega\ t/2$).
The exact solution is known to be the straight-line $V^1(s) = V^2(s) = 0, V^3(s) = \alpha t$ resulting in the complexity \cite{Haque:2021hyw}
\begin{equation}
	\mathcal{C}_{\rm su(1,1)}^{(\rm exact)}=\min_{n} |2\pi n-\alpha t|\, ,
 \label{eq:su11exactC}
\end{equation}
The global minimization in (\ref{eq:su11exactC}) takes into account the periodicity of operator space in this direction, resulting in the sawtooth-like behavior (see \cite{Haque:2021hyw} for more discussion).
Let's compare this to the perturbative result. With these choices of gate set and target operator, the sub-leading contribution to the universal perturbative result (\ref{C22}) vanishes $\Omega_2^I = 0$, so the perturbative complexity grows linearly with time
\be
{\mathcal C}_{\rm su(1,1)}^{(\rm pert)} = |\omega| t = \alpha t\, ,
\ee
matching the exact solution (\ref{eq:su11exactC}) at early times (the perturbative result cannot easily see the periodicity effect described in \cite{Haque:2021hyw}). 

More generally, for an arbitrary target operator as parameterized by (\ref{eq:SU11Target}) and an equal-cost metric, the perturbative result (\ref{C22}) leads to the behavior
\be
{\mathcal C}_{\rm su(1,1)}^{(\rm pert)} &=& |\omega| t \left[1-\frac{1}{6} \frac{((\omega^1)^2 + (\omega^2)^2) (\omega^3)^2}{|\omega|^2}\ t^2\right]
\nonumber \\
&=& 2 \sqrt{r(t)^2 + \theta(t)^2} \left[1-\frac{2}{3} \frac{r(t)^2 \theta(t)^2}{r(t)^2 + \theta(t)^2}\right]\, ,
\ee
where in the second line we re-expressed the result in terms of the time-dependent squeezing and rotation angle parameters. This matches the corresponding exact results of \cite{Haque:2021hyw} for $r \ll 1$.

The advantage of the perturbative result (\ref{C22}) is that the early-time complexity can easily be obtained in cases where the exact result may be difficult to find analytically. For example, consider a weighting scheme where the $\hat e_1, \hat e_2$ gates, corresponding to an inverted oscillator, are weighted more than the free Hamiltonian gate $\hat e_3$, e.g.
\be
G_{IJ} = \begin{pmatrix}
\mu & 0 & 0 \cr
0 & \mu & 0 \cr
0 & 0 & 1
\end{pmatrix}\, ,
\ee
for $\mu \gg 1$. 
As discussed in Section \ref{subsec:PertComplexity}, the \emph{generic} expectation of the complexity for the case when some of the gates have a large weight factor is that the complexity initially grows at a rapid linear rate proportional to $\sqrt{\mu}\, t$, but the higher order terms quickly become important around the timescale
(\ref{eq:tpert}) $t_{\rm pert} \sim |\omega|^{-1} \sim 1/(\sqrt{\mu} |\omega_{\rm hard}|)$, which is suppressed parametrically by $\mu$, shutting off the rapid linear growth.
For the $\mathfrak{su}(1,1)$ gates, the corresponding complexity for an arbitrary target operator takes the form
\be
{\mathcal C}_{\rm \mu, su(1,1)}^{({\rm pert})} \approx \sqrt{(\omega^1)^2+(\omega^2)^2} \sqrt{\mu}\ t \left[1 - \frac{1}{24} (\omega^3)^2\  t^2\right]\, .
\ee
As expected, the leading-order behavior is rapid linear growth proportional to $\sqrt{\mu}\, t$; however, because of the relations between the structure constants, the higher order terms only become important at the parametrically longer timescale set by the periodicity in the $\omega^3$ direction, $t_{\rm conj} \sim (\omega_3)^{-1}$.
Thus, the rapid linear growth occurs for a parametrically longer time, a feature that appears to be unique to the $\hat e_1, \hat e_2$ gates of $\mathfrak{su}(1,1)$ due to their relationship to an inverted oscillator.

\subsection{Coupled Oscillators}
\label{sec:coupled}

Let us extend the single harmonic oscillator of the previous section to a pair of coupled oscillators with Hamiltonian controlled by a coupling parameter $\beta$
\begin{align}
\label{eq:coupledH}
	\hat H_{\rm coupled} &= \frac{1}{2m} \left(\hat p_1^2 + \hat p_2^2 \right) + \frac{m\alpha^2}{2} \left(\hat x_1^2 + \hat x_2^2\right)+\beta \left(\hat x_1 \hat x_2\right)\\
 &= \alpha (\hat a_1^\dagger \hat a_1 + \hat a_2^\dagger \hat a_2)+\tfrac{\tilde\beta}{2} (\hat a_1 \hat a_2+\hat a_1\hat a_2^{\dagger} + \hat a_2 \hat a_1^{\dagger}+ \hat a_1^{\dagger} \hat a_2^{\dagger})\, , \nonumber
\end{align}
where $\tilde \beta \equiv \beta/(m\alpha)$, we re-wrote the Hamiltonian in terms of the (free) raising and lowering operators
\be
\hat a&=&\sqrt{\frac{m\alpha}{2}}\left({\hat {x}}+{\frac{i}{m\alpha }}{\hat {p}}\right)\, , \hspace{.4in} \hat a^{\dagger }={\sqrt{\frac{m\alpha}{2}}}\left({\hat {x}}-{\frac{i}{m\alpha}}{\hat {p}}\right)\,,
\label{eq:raisingLoweringDef}
\ee
and we assumed the two oscillators have the same fundamental frequency $\alpha$.
The Hamiltonian (\ref{eq:coupledH}) can also be seen as arising from the coupled oscillators,
\be
\label{eq:coupledH2}
\hat H = \frac{1}{2m} \left(\hat p_1^2+\hat p_2^2\right) + \frac{m k^2}{2}\left(\hat x_1^2 + \hat x_2^2\right) + \frac{\beta}{2}\left(\hat x_1-\hat x_2\right)^2\, ,
\ee
with fundamental frequency $\alpha^2 = k^2 + \beta/m$.
The relative strength of the coupling between the oscillators compared to their fundamental frequency is controlled by the dimensionless parameter $\tilde\beta/\alpha$.

The quantum circuit complexity of two coupled oscillators was studied using the wavefunction method in \cite{Bhattacharyya:2020iic} in the context of open systems.
The operators in (\ref{eq:coupledH}) are part of the $\mathfrak{sp}\left(4,\mathbb{R}\right)$ algebra, and we will take the corresponding 10 generators as our fundamental gates:
\begin{align}
	& \hat e_0=\tfrac12 ( \hat a_1\hat  a_1^{\dagger}+ \hat a_2 \hat a_2^{\dagger}),	&& \hat e_1=\tfrac12 ( \hat a_1 \hat a_1^{\dagger}- \hat a_2 \hat a_2^{\dagger},) \nonumber\\
	&\hat e_2=\tfrac12 (\hat a_1^{\dagger}\hat a_2^{\dagger}+\hat a_1\hat a_2),  &&\hat e_3=\tfrac{i}2 (\hat a_1^{\dagger}\hat a_2^{\dagger}-\hat a_1\hat a_2),\nonumber\\
\label{eq:coupledGates}		&\hat e_4=\tfrac12 ( \hat a_2 \hat a_1^{\dagger}+\hat a_1 \hat a_2^{\dagger}),  &&\hat e_5=\tfrac{i}2 (\hat a_2 \hat a_1^{\dagger}-\hat a_1 \hat a_2^{\dagger}),\\
				&\hat e_6=\tfrac12 ( \hat a_1^{\dagger2}+\hat a_1^2),  &&\hat e_7=\tfrac{i}2 ( \hat a_1^{\dagger2}-\hat a_1^2),\nonumber\\
					&\hat e_8=\tfrac12 ( \hat a_2^{\dagger2}+\hat a_2^2),  &&\hat e_9=\tfrac{i}2 ( \hat a_2^{\dagger2}-\hat a_2^2),\nonumber
\end{align}
In terms of these gates, the coupled Hamiltonian (\ref{eq:coupledH}) can be written
\be
\label{eq:coupledHgates}
\hat H_{\rm coupled} = 2\alpha \hat e_0 + \tilde \beta \left(\hat e_2 + \hat e_4\right)\, .
\ee
This algebra is closed, with the commutation relations
\begin{align}
&[\hat{e}_0,\hat{e}_2]=-i
   \hat{e}_3,[\hat{e}_0,\hat{e}_3]=i
   \hat{e}_2,[\hat{e}_0,\hat{e}_6]=-i
   \hat{e}_7,[\hat{e}_0,\hat{e}_7]=i
   \hat{e}_6,[\hat{e}_0,\hat{e}_8]=-i
   \hat{e}_9,[\hat{e}_0,\hat{e}_9]=i \hat{e}_8\nonumber\\
   &[\hat{e}_1,\hat{e}_4]=-i
   \hat{e}_5,[\hat{e}_1,\hat{e}_5]=i
   \hat{e}_4,[\hat{e}_1,\hat{e}_6]=-i
   \hat{e}_7,[\hat{e}_1,\hat{e}_7]=i
   \hat{e}_6,[\hat{e}_1,\hat{e}_8]=i
   \hat{e}_9,[\hat{e}_1,\hat{e}_9]=-i
   \hat{e}_8\nonumber\\&[\hat{e}_2,\hat{e}_3]=i
   (\hat{e}_0-\tfrac{1}{2}),[\hat{e}_2,\hat{e}_4]=\tfrac{i}{2}
   (\hat{e}_7+\hat{e}_9),[\hat{e}_2,\hat{e}_5]=\tfrac{i}{2}
   (\hat{e}_8-\hat{e}_6),[\hat{e}_2,\hat{e}_6]=-i \hat{e}_5,[\hat{e}_2,\hat{e}_7]=i
   \hat{e}_4\nonumber\\&[\hat{e}_2,\hat{e}_8]=i
   \hat{e}_5,[\hat{e}_2,\hat{e}_9]=i
   \hat{e}_4\nonumber\\&[\hat{e}_3,\hat{e}_4]=-\tfrac{i}{2}
   (\hat{e}_6+\hat{e}_8),[\hat{e}_3,\hat{e}_5]=\tfrac{i}{2}
   (\hat{e}_9-\hat{e}_7),[\hat{e}_3,\hat{e}_6]=-i \hat{e}_4,[\hat{e}_3,\hat{e}_7]=-i
   \hat{e}_5,[\hat{e}_3,\hat{e}_8]=-i
   \hat{e}_4,[\hat{e}_3,\hat{e}_9]=i
   \hat{e}_5\nonumber\\&[\hat{e}_4,\hat{e}_5]=-i \hat{e}_1,[\hat{e}_4,\hat{e}_6]=-i
   \hat{e}_3,[\hat{e}_4,\hat{e}_7]=i
   \hat{e}_2,[\hat{e}_4,\hat{e}_8]=-i
   \hat{e}_3,[\hat{e}_4,\hat{e}_9]=i
   \hat{e}_2\nonumber\\&[\hat{e}_5,\hat{e}_6]=-i
   \hat{e}_2,[\hat{e}_5,\hat{e}_7]=-i
   \hat{e}_3,[\hat{e}_5,\hat{e}_8]=i
   \hat{e}_2,[\hat{e}_5,\hat{e}_9]=i
   \hat{e}_3\nonumber\\&[\hat{e}_6,\hat{e}_7]=i (2 \hat{e}_0+2
   \hat{e}_1-1),[\hat{e}_8,\hat{e}_9]=i (2 \hat{e}_0-2
   \hat{e}_1-1)\,.
\end{align}

Taking the perturbative-in-time result for complexity (\ref{C22}) and an equal-cost metric $G_{IJ} = \delta_{IJ}$, the leading-order linear in time behavior for the target operator $\hat{\mathcal U}_{\rm target} = {\rm exp}\left[-i t \hat H_{\rm coupled}\right] = {\rm exp}\left[-i t \omega^I \hat e_I\right]$ is
\be
{\mathcal C}_{\rm coupled} \approx |\omega| t = \left(4 \alpha^2 + 2 \tilde \beta^2\right)^{1/2}\, t\, .
\ee
In the decoupling limit $\tilde \beta \rightarrow 0$, this indeed gives the correct scaling for two decoupled oscillators ${\mathcal C}\rightarrow 2\alpha\, t$.
At next-to-leading order in time, the corrections depend on the magnitude of the vector
\be
\Omega_2^I = G^{IL} f_{LK}^J \omega^K G_{JM} \omega^M\, .
\label{eq:coupledOmega2}
\ee
Using the structure constants extracted from the algebra above, it is straightforward to find $|\Omega_2|^2 = 16\alpha^2 \tilde \beta^2 + 8 \tilde \beta^4$
so that the next-to-leading order complexity takes the simple form
\be
\label{eq:coupledComplexity}
{\mathcal C}_{\rm coupled} \approx |\omega| t \left(1-\frac{1}{24} \frac{|\Omega_2|^2}{|\omega|^2}\, t^2\right) = \left(4 \alpha^2 + 2 \tilde \beta^2\right)^{1/2}\, t \left(1-\frac{1}{6}\, \tilde \beta^2 t^2\right)\, .
\ee
We see that while the growth rate of complexity is set by the prefactor $|\omega| = \sqrt{4\alpha^2 + 2 \tilde \beta^2}$, the timescale for the corrections in (\ref{eq:coupledComplexity}) to be important is set only by the coupling between the oscillators $\tilde\beta$. For weak coupling $\tilde \beta \ll \alpha$, these two timescales are parametrically separate from each other and the linear growth of complexity is dominant.
However, for strong coupling $\tilde \beta \gg \alpha$ the rapid linear growth of complexity with rate $\tilde \beta^{-1}$ is quickly shut off by the interactions within the same timescale $t \sim \tilde \beta^{-1}$, so that only ${\mathcal O}(1)$ growth in complexity can occur before the next-order-in-time corrections become important. 
In the next section, we will see that this is true more generally for interacting systems: in the strong coupling limit, a rapid linear rise in complexity receives corrections to this linear growth after only ${\mathcal O}(1)$ growth.
While it is possible that these higher order corrections shut off complexity, it is also possible that they lead to a different regime of linear growth of complexity with time. Unfortunately, our perturbative-in-time analysis does not allow us to determine this intermediate- to late-time behavior.

It is also possible to rewrite the coupled oscillators (\ref{eq:coupledH2}) in terms of the normal modes $\hat x_{\pm} = (\hat x_1\pm \hat x_2)/\sqrt{2}, \hat p_\pm = (\hat p_1\pm \hat p_2)/\sqrt{2}$
\be
\hat H_{\rm normal} = \frac{1}{2m} \left(\hat p_+^2 + \hat p_-^2\right) + \frac{m}{2} \left(\alpha_+^2\hat x_+^2 + \alpha_-^2 \hat x_-^2\right) = \alpha_+ \hat a_+^\dagger \hat a_+ + \alpha_- \hat a_-^\dagger \hat a_-\, ,
\ee
where $\alpha_+^2 = k^2, \alpha_-^2=k^2+2\beta/m$.
In this normal-mode space, the two oscillators decouple into two independent, free harmonic oscillators, so that the complexity of the target time-evolution operator $\hat {\mathcal U}_{\rm normal} = {\rm exp}\left[-it\hat H_{\rm normal}\right]$ grows strictly linearly with time according to Section \ref{sec:HarmonicOscillator}
\be
{\mathcal C}_{\rm normal mode} = |\omega|t = \sqrt{\alpha_+^2+\alpha_-^2}\ t = \sqrt{2}\,\alpha\, t\, ,
\ee
in terms of the fundamental frequency $\alpha$.
At first glance this seems to contradict the result (\ref{eq:coupledComplexity}) using the original $\hat x_1, \hat x_2$ coordinates, in which the coupling between the oscillators causes the complexity to deviate from a linear dependence on time.
However, the different behavior in complexity is merely reflecting the difference in the set of gates used to build the target unitary. The original coordinates used the set (\ref{eq:coupledGates}), which have non-trivial commutation relations. In contrast, the normal mode gate set $\hat a_{\pm}^\dagger\hat a_\pm$ are combinations of the gates in (\ref{eq:coupledGates}) and decouple from each other.
The ``right'' gate set depends on the details of the situation of interest (perhaps the physical setup or interaction with an external system) which determines the natural set of gates that are used to build the target unitary. Alternatively, a more fundamental approach might be to minimize the complexity over all choices of the gate set, in which case the gates built from the original coordinates (\ref{eq:coupledGates}) may be a better choice than the normal mode gates during some range of time.

\subsection{Anharmonic Oscillator}
\label{sec:AnharmonicOscillator}

The perturbative framework of Section \ref{sec:pert} can also be used extract the behavior of the complexity for circuits that are non-Gaussian and may be difficult to study using wavefunction-based approaches to circuit complexity or constructions with generalized coherent-states.
Circuits of this type are of interest as they encode the effects of non-linearities and interactions, and there are few, if any, approaches that can be used to calculate their corresponding circuit complexity.

As a simple representative example, consider the \emph{anharmonic} oscillator Hamiltonian with fundamental frequency $\alpha$
\be
\hat H_{\rm anharm} &=& \frac{1}{2m} \hat p^2 + \frac{m\alpha^2}{2} \hat x^2 + \lambda \hat x^4 \nonumber \\
&=& \left(\alpha + 3 \tilde \lambda \right) \hat e_0 + \frac{1}{2}\tilde \lambda \left(3 \hat e_1 + 3 \hat e_2 + 2 \hat e_3 + \hat e_4\right)\, ,
\label{eq:AnHarmonicH}
\ee
where $\tilde \lambda \equiv \lambda/(m^2\alpha^2)$, we re-wrote the Hamiltonian in terms of the (free) raising and lowering operators
\be
\hat a&=&\sqrt{\frac{m\alpha}{2}}\left({\hat {x}}+{\frac{i}{m\omega }}{\hat {p}}\right)\, , \hspace{.4in} \hat a^{\dagger }={\sqrt{\frac{m\alpha}{2}}}\left({\hat {x}}-{\frac{i}{m\omega}}{\hat {p}}\right)\,,
\ee
and used the ``target'' gates $\{\hat e_0\ldots \hat e_4\}$
\begin{eqnarray}
\label{eq:AnHarmonicTargetGates}
    \hat e_0 &=& \hat a^\dagger \hat a\, ,  \\
    \hat e_1 &=&\left(\hat a^{\dagger 2}+\hat a^2\right)\, ,\nonumber \\
    \hat e_2 &=& \hat a^{\dagger 2} \hat a^2\, , \nonumber \\
    \hat e_3 &=&\left(\hat a^{\dagger 3} \hat a+\hat a^{\dagger} \hat a^3\right)\, ,\nonumber \\
    \hat e_4 &=& \left(\hat a^{\dagger 4}+\hat a^4\right)\,.\nonumber
\end{eqnarray}
As with the coupled oscillators in the previous section, we see that the strength of the anharmonic interaction relative to the usual harmonic oscillator is controlled by the dimensionless parameter $\tilde \lambda/\alpha$.

The algebra of these ``target'' gates is not closed by itself; for example, the commutator of two of these gates generates
\be
\left[\hat e_1,\hat e_3\right] = -2\left(\hat a^{\dagger 4}-\hat a^4\right)\, ,
\label{eq:targetGeneratedGate}
\ee
which is not in the set (\ref{eq:AnHarmonicTargetGates}). 
We could try and expand our gate set to include the new operator appearing in (\ref{eq:targetGeneratedGate}) as $\hat e_7 = i\left(\hat a^{\dagger 4}-\hat a^4\right)$ (the numbering anticipates an ordering later), but then we see that the algebra of operators of this expanded gate set generates yet another new gate, now higher order in products of $\hat a^\dagger, \hat a$, as in
\be
\left[\hat e_2,\hat e_7\right] = 12 i\hat e_4 + 8i\left(\hat a^{\dagger 5}\hat a + \hat a^\dagger \hat a^5\right)\, .
\ee
It is straightforward to see that this process will not stop, with increasingly higher powers of $\hat a^\dagger, \hat a$ generated, so that the number of generated operators constructed from (\ref{eq:AnHarmonicTargetGates}) in this way is \emph{infinite}.
At first glance, this seems to imply that it is not possible to calculate the complexity of a unitary operator constructed from the anharmonic oscillator gates (\ref{eq:AnHarmonicTargetGates}) without working with an infinite number of fundamental gates, { and a corresponding infinite-dimensional manifold of the space of gates}.

However, we will take advantage of the fact that the perturbative complexity (\ref{C22}) up to some finite order in $t$ only requires the target gates (\ref{eq:AnHarmonicTargetGates}) and some \emph{finite set} of generated gates\footnote{
 Since we will only need a finite number of gates at any order in perturbation theory to construct the target operator, the corresponding group sub-manifold generated by these gates is finite-dimensional and the minimization of the circuit depth through the Euler-Arnold equation is well-defined on this sub-manifold. At late times, however, the required group manifold appears to be infinite-dimensional when parameterized with this gate set, and more care may be needed in the Euler-Arnold minimization procedure. The late-time behavior of the complexity for the anharmonic oscillator, however, is beyond the scope of this work.}.
For example, consider a target unitary operator of the form
\be
\hat {\mathcal U}_{\rm target} = e^{-i \hat H_{\rm anharm}t} = e^{-i t\omega^I \hat e_I}\, ,
\ee
where $\hat H_{\rm anharm}$ and the target operator vector components $\omega^I$ can be extracted from (\ref{eq:AnHarmonicH}).
For a equal-cost metric $G_{IJ} = \delta_{IJ}$, the \emph{leading-order} result is simply given by
\be
{\mathcal C}\approx |\omega| t = \sqrt{\alpha^2+\frac{89}{16}\tilde \lambda^2}\, t\, ,
\ee
and does not require any additional gates beyond the target gates (\ref{eq:AnHarmonicTargetGates}). It is also important to note that this result does not require the anharmonic parameter $\tilde \lambda$ to be small compared to the fundamental frequency $\alpha$, so the result is perturbative in time, but is valid for all $\tilde \lambda$, even for $\tilde \lambda \gg \alpha$.

At the next order in time, the complexity
\be
{\mathcal C}\approx |\omega| t \left(1-\frac{1}{24} \frac{|\Omega_2|^2}{|\omega|^2}\, t^2\right)\, ,
\label{CAnHarmonic}
\ee
becomes sensitive to the algebra of the gate set through 
\be
\Omega_2^I = G^{IL} f_{LK}^J \omega^K G_{JM} \omega^M\, .
\label{AnHarmonicOmega2}
\ee
In order to analyze this contribution, it helps to make a distinction between the ``target gate'' set (\ref{eq:AnHarmonicTargetGates}), which we will now denote as $\{\hat e_i\} = \{\hat e_0, \hat e_1, \hat e_2, \hat e_3, \hat e_4\}$ with latin indices $i,j,...$, and a set of ``generated gates'' denoted as $\{\hat e_\alpha\}$ with greek indices $\alpha, \beta, ...$.
By definition the target operator vector components are zero along the ``generated'' directions, $\{\omega^I\} =\{\omega^i,\omega^\alpha\} = \{\omega^i,0\}$, but the vector $\Omega_2^I$ may in principle have components along the target gate directions $\Omega_2^i$ as well as the generated gate directions $\Omega_2^\alpha$.
Examining the structure of (\ref{AnHarmonicOmega2}), first notice that components of $\Omega_2^I$ along the target gate directions only contain contributions from structure constants among the target gates themselves,
\be
\Omega_2^i = G^{i\ell} f_{\ell K}^J \omega^K G_{JM} \omega^M = G^{i\ell} f_{\ell k}^j \omega^k G_{jm} \omega^m\, ,
\ee
where we will assume that the cost metric $G_{IJ}$ does not mix the target and generated gates.
The algebra among the target gates is known, and can easily be calculated.
Along the generated gate directions, the vector $\Omega_2^\alpha$ takes the form
\be
\Omega_2^\alpha = G^{\alpha \beta} f_{\beta K}^J \omega^K G_{JM} \omega^M = G^{\alpha\beta} f_{\beta k}^j \omega^k G_{jm} \omega^m\, .
\label{eq:AnHarmonicOmega2}
\ee
These components will be non-zero if there are non-vanishing structure constants of the form $f_{\beta k}^j$ mixing target and generated gates, which arise through an algebra with contributions of the form
\be
\left[\hat e_\beta,\hat e_k\right] = i f_{\beta k}^j \hat e_j + \ldots
\label{eq:AnHarmonicAlgTerm}
\ee
(where $...$ could be contributions with a different form of the structure constants).
The task of calculating the next-to-leading order corrections to the complexity (\ref{CAnHarmonic}) can now be reduced to finding the set of generated gates $\{\hat e_\alpha\}$ that lead to an algebra of the form (\ref{eq:AnHarmonicAlgTerm}), find the corresponding structure constants $f_{\beta k}^j$, use these to calculate the vector $\Omega_2^\alpha$ in (\ref{eq:AnHarmonicOmega2}), which then directly leads to the complexity (\ref{CAnHarmonic}).

The details of the algebra and resulting components of $\Omega_2^I$ for the anharmonic oscillator can be found in Appendix \ref{app:Anharmonic}.
Here we will summarize the result.
For the target gates (\ref{eq:AnHarmonicTargetGates}), there are a finite set of generated gates $\hat e_\alpha$ that will generate an algebra of this form.
The generated gates that can potentially contribute to terms of the form (\ref{eq:AnHarmonicAlgTerm}) are
\begin{align}
\label{AnHarmonicGeneratedGates}
    \hat e_5 & = i\left(\hat a^{\dagger 2} - \hat a^2\right),        & \hat e_8 &= \hat a^{\dagger 6}+\hat a^6\, , \\
    \hat e_6 & = i\left(\hat a^{\dagger 3}\hat a - \hat a^{\dagger} \hat a^3\right),   &      \hat e_9 &= i\left(\hat a^{\dagger 6}-\hat a^6\right), \nonumber \\
    \hat e_7 &= i\left(\hat a^{\dagger 4} - \hat a^4\right),      &    \hat e_{10} &= \hat a^{\dagger 5}\hat a + \hat a^{\dagger} \hat a^5\, , \nonumber \\
    &   &   \hat e_{11} &= i\left(\hat a^{\dagger 5}\hat a - \hat a^{\dagger} \hat a^5\right),\nonumber  \\
    &   &   \hat e_{12} &= \hat a^{\dagger 4}\hat a^2 + \hat a^{\dagger 2}\hat a^4\, , \nonumber \\
    &   &   \hat e_{13} &= i\left(\hat a^{\dagger 4}\hat a^2 - \hat a^{\dagger 2}\hat a^4\right), \nonumber \\
    &   &   \hat e_{14} &= \hat a^{\dagger 3}\hat a^3\, .\nonumber 
\end{align}
By explicitly calculating the commutation relations between the target gates (\ref{eq:AnHarmonicTargetGates}) and the generated gates (\ref{AnHarmonicGeneratedGates}) we can extract the $f_{\beta k}^j$ for (\ref{eq:AnHarmonicAlgTerm}) and calculate (\ref{eq:AnHarmonicOmega2}) for a given choice of cost metric $G_{IJ}$. Interestingly, the components of $\Omega_2^i$ along the target gate directions vanish identically, so that only the generated gate components $\Omega_2^\alpha$ contribute to the next-to-leading order in time result for complexity.
For an equal-cost metric $G_{IJ} = \delta_{IJ}$, we find
\be
\label{eq:AnHarmonicOmega2Result}
|\Omega_2|^2 = \tilde \lambda^2 \left(a\,\alpha^2 + b\, \alpha\, \tilde \lambda + c\, \tilde \lambda^2\right)\, ,
\ee
where $a,b,c$ are pure numbers with values $a = 24\, 749\, , b = 287\, 653\, , c = 1\, 257\, 067$.
The resulting complexity (\ref{CAnHarmonic}) thus becomes
\be
{\mathcal C}_{\rm anharm} &\approx & |\omega| t \left(1-\frac{1}{24} \frac{|\Omega_2|^2}{|\omega|^2}\, t^2\right) \nonumber \\
&=& \left(\alpha^2 + \frac{89}{16} \tilde \lambda^2 \right)^{1/2}\, t \left(1-\frac{1}{24} \frac{\tilde \lambda^2 \left(a\,\alpha^2 + b\, \alpha\, \tilde \lambda + c\, \tilde\lambda^2\right)}{\alpha^2 + \frac{89}{16} \tilde \lambda^2}\, t^2\right)\, .
\label{eq:AnHarmonicComplexityResult}
\ee
An interesting feature of this calculation is that it is not necessary to assume that the anharmonic parameter $\tilde \lambda$ is small (in this case, small in comparison to the fundamental frequency $\alpha$ of the corresponding harmonic oscillator).
Instead, the result (\ref{eq:AnHarmonicComplexityResult}) is valid for any $\tilde \lambda$.
While the complexity for the anharmonic oscillator follows the general features of the result (\ref{C22}) -- namely that the early-time behavior of complexity is linear in time, and next-to-leading-order in time corrections to this behavior have a fixed negative sign -- the result (\ref{eq:AnHarmonicComplexityResult}) is noteworthy primarily because it provides a mechanism to calculate the complexity of an anharmonic oscillator in closed form, without approximation or a perturbative series in $\tilde \lambda$.
In particular, from (\ref{eq:AnHarmonicComplexityResult}) we see that in the limit of a weak anharmonic parameter $\tilde \lambda \ll \alpha$, the linear growth rate of complexity is controlled by the fundamental frequency $\alpha$, while the timescale for corrections to this linear growth are controlled by the anharmonic parameter $\tilde \lambda$,
${\mathcal C}\approx \alpha t (1-a\tilde \lambda^2 t^2/24)$.
Alternatively, in the limit of a strong anharmonic parameter $\tilde \lambda \gg \alpha$, both of these scales are controlled by $\tilde \lambda^{-1}$ since ${\mathcal C} \approx \tilde \lambda\, t (1-c \tilde \lambda^2 t^2/24)$.
Mirroring what we saw in Section \ref{sec:coupled} for the complexity of coupled oscillators, in the case of strong coupling the rapid linear growth in complexity is proportional to the coupling, but the linear approximation breaks down with in the same timescale so that only ${\mathcal O}(1)$ growth in complexity can occur before the linear perturbative approximation fails.

It is straightforward to extend the analysis above to calculate complexity to higher-order in time beyond ${\mathcal O}(t^3)$, using the results of Section \ref{sec:PertSolnsN} and the reasoning employed here.
At higher orders in time, the complexity will depend on higher powers of the structure constants contracted together, which in turn will require more generated gates to be added to the set (\ref{AnHarmonicGeneratedGates}); however, at any given order in time, the number of generated gates (and corresponding algebra and structure constants) will remain finite.

{
Finally, we comment briefly on the effect of different weighting schemes for the gates.}
One potential weighting scheme is to weight the ``generated'' gates by a cost factor $\mu$ compared to the target gates. Since the components $\Omega_2^i$ vanish along the target gate directions, the cost factor shows up as an inverse power of $\mu$ in $\Omega_2^\alpha$ from (\ref{eq:AnHarmonicOmega2}):
\be
{\mathcal C} &\approx & \left(\alpha^2 + \frac{89}{16} \tilde \lambda^2 \right)^{1/2}\, t \left(1-\frac{1}{24} \frac{\tilde \lambda^2 \left(a\,\alpha^2 + b\, \alpha\, \tilde \lambda + c\, \tilde\lambda^2\right)}{\mu(\alpha^2 + \frac{89}{16} \tilde \lambda^2)}\, t^2\right)\, .
\ee
Since none of the ``generated'' gates have a contribution to $\omega^I$, when the cost factor is large $\mu \gg 1$ the higher order corrections to complexity vanish and complexity simply becomes linear in time and independent of $\mu$, indicating that the generated gates become projected out, as expected.
{
Another weighting scheme is to assign a cost for each gate associated with the corresponding powers of the raising and lowering operators. Under this weighting scheme, gates $\hat e_0, \hat e_1$ are second order in $\hat a, \hat a^\dagger$ and would receive no cost; gates $\hat e_2...\hat e_4$ are fourth order in $\hat a, \hat a^\dagger$, and are assigned a cost $\mu$ associated with some potential difficulty in applying multiple raising and lowering operators; finally, gates $\hat e_8...\hat e_{14}$ are assigned a cost $\mu^2$ since they are sixth order in $\hat a, \hat a^\dagger$ (and are therefore presumably more difficult to implement).
The resulting complexity for $\mu \gg 1$ takes the generic form
\be
{\mathcal C} \approx \sqrt{6\mu}\, \tilde \lambda\, t \left(1-d \tilde\lambda^2\, t^2\right)\, ,
\ee
for some constant $d = 149629/96$.
In general, it is not clear what, if any, cost metric makes sense. In general, the results are simply a variation on the discussion of cost metrics of Section \ref{subsec:PertComplexity} around (\ref{Veasy}), and we see similar qualitative results: the linear growth in complexity is dominated by the ``hard'' directions and is proportional to $\sqrt{\mu}$. Whether, and how, the sub-leading behavior is also controlled by the cost factor $\mu$ depends on the specific structure constants of the gates and the weighting scheme.
}

\subsection{Free and Interacting Scalar Fields}
\label{sec:scalarfield}

The previous subsections can be viewed as a warmup for the analysis of free and interacting scalar fields on a lattice.
Consider first a free massive scalar field in $d$ spacetime dimensions with Hamiltonian
\be
\label{eq:ScalarFieldH}
\hat H = \frac{1}{2} \int d^{d-1}x\left[\hat \pi^2(x) + \left(\vec{\nabla}\hat \phi\right)^2 + m^2 \hat \phi^2\right]\, ,
\ee
on a lattice with spacing $\delta$. We can rewrite (\ref{eq:ScalarFieldH}) as a sum over the lattice sites $\vec{n}$
\be
\label{eq:Hlattice}
\hat H_{\rm lattice} = \sum_{\vec{n}} \left[\frac{\hat P(\vec{n})^2}{2M} + \frac{M}{2} k^2 \hat X(\vec{n})^2 + \frac{\beta}{2} \sum_i (\hat X(\vec{n}) - \hat X(\vec{n}-\hat x_i))^2\right]\, ,
\ee
where the $\hat x_i$ are unit vectors that point along the lattice and we defined $\beta = 1/\delta^3$, $M = 1/\delta$, $k = m$, and $\hat X = \delta^{d/2}\hat\phi$, $\hat P = \delta^{d/2}\hat\pi$.

To be specific, let's consider a 1-dimensional chain of $N$ oscillators in $d = 2$ coupled as in (\ref{eq:Hlattice})
\be
\label{eq:HFreelattice2}
\hat H_{\rm chain} = \sum_{i=1}^N \left[\frac{\hat P_i^2}{2M} + \frac{M\alpha^2}{2} \hat X_i^2 + \frac{\beta}{2} \left(\hat X_i - \hat X_{i+1}\right)^2\right]\, ,
\ee
where we take the boundary conditions to be periodic, e.g.~$\hat X_{N+1} = \hat X_1$.
The analogy with Section \ref{sec:coupled}, in particular (\ref{eq:coupledH2}) is clear, so that the chain can be considered as $N$ harmonic oscillators with nearest-neighbor coupling.
After substituting in expressions for $\hat X_i,\hat P_i$ in terms of raising and lowering operators along the lines of (\ref{eq:raisingLoweringDef}), the lattice Hamiltonian (\ref{eq:HFreelattice2}) can be written as
\be
\label{eq:Hlattice2}
\hat H_{\rm chain} = \sum_{i=1}^N \left[\alpha \hat a_i^\dagger \hat a_i + \frac{\tilde \beta}{2} \left(\hat a_i^\dagger \hat a_{i+1}^\dagger + \hat a_i \hat a_{i+1} + \hat a_i^\dagger \hat a_{i+1} + \hat a_i \hat a_{i+1}^\dagger\right)\right]\, ,
\ee
where the fundamental frequency is $\alpha = \sqrt{m^2 + \beta/M} = \sqrt{m^2 + 1/\delta^2}$ and $\tilde \beta = \beta/(M\alpha)$, as with the coupled oscillators from Section \ref{sec:coupled}.
The strength of the coupling between the oscillators is controlled by the dimensionless quantity
\be
\label{eq:coupledStrength}
\frac{\tilde \beta}{\alpha} = \frac{\beta}{M\alpha^2} = \frac{1}{m^2\delta^2 + 1} \approx \begin{cases}
    \frac{1}{m^2\delta^2} \ll 1 & \mbox{ for } m \gg 1/\delta \cr
    {\mathcal O}(1) & \mbox{ for } m \ll 1/\delta\hspace{.2in} (\mbox{i.e. }\delta \rightarrow 0)\,.
    \end{cases}
\ee
Thus we see that the ``continuum limit'' in which the lattice spacing goes to zero $\delta \rightarrow 0$ leads to an ${\mathcal O}(1)$ coupling between nearest neighbors in the corresponding coupled oscillator system; since the analysis of coupled oscillators in Section \ref{sec:coupled} does not rely on small coupling $\tilde \beta$, we can probe the continuum limit regime with our analysis.

In analogy with the pair of coupled oscillators, it is natural to choose gates that are linear combinations of generators of $\mathfrak{sp}\left(2N,\mathbb{R}\right)$
\begin{align}
\hat b_i^+ &= \frac{1}{2} \left(\hat a_i^\dagger \hat a_{i+1}^\dagger + \hat a_i \hat a_{i+1}\right), & \hat b_i^- &= \frac{i}{2} \left(\hat a_i^\dagger \hat a_{i+1}^\dagger - \hat a_i \hat a_{i+1}\right), \nonumber \\
\hat c_i^+ &= \frac{1}{2} \left(\hat a_i^\dagger \hat a_{i+1} + \hat a_i \hat a_{i+1}^\dagger\right), & \hat c_i^- &= \frac{i}{2} \left(\hat a_i^\dagger \hat a_{i+1} - \hat a_i \hat a_{i+1}^\dagger\right), \\
\hat d_i^+ &= \frac{1}{2} \left(\hat a_i^{\dagger 2} + \hat a_i^2\right), & \hat d_i^- &= \frac{i}{2} \left(\hat a_i^{\dagger 2} - \hat a_i^2\right), \nonumber \\
\hat h_i & = \hat a_i^\dagger \hat a_i\, . & & \nonumber
\end{align}
With these gates, (\ref{eq:Hlattice2}) becomes
\be
\hat H_{\rm chain} = \sum_{i=1}^N \left[\alpha \hat h_i^+ + \frac{\beta}{M\alpha} \left(\hat b_i^+ + \hat c_i^+\right)\right]\, ,
\ee
which we recognize as $N$-copies of the coupled oscillator of Section \ref{sec:coupled}. Therefore, for the target operator $\hat {\mathcal U}_{\rm target} = {\rm exp}\left[-it\hat H_{\rm chain}\right]$, the complexity at early times is
\begin{align}
\label{eq:complexityChain1}
{\mathcal C}_{\rm chain} &= \sqrt{N}\sqrt{4\alpha^2 + 2 \left(\frac{\beta}{M\alpha}\right)^2}\,t\, \left(1-\frac{1}{6} \left(\frac{\beta}{M\alpha}\right)^2\right) \\
\label{eq:complexityChain2}
& = \sqrt{N}\sqrt{4m^2 + \frac{4}{\delta^2} +2 \frac{1/\delta^4}{m^2+1/\delta^2}}\, t\, \left(1-\frac{1}{24} \frac{1/\delta^4}{m^2+1/\delta^2}\, t^2\right) \\
\label{eq:complexityChain3}
& \approx \begin{cases}
    2\sqrt{N}\, m\, t \left(1-\frac{1}{24} \frac{1}{m^2 \delta^2} \frac{t^2}{\delta^2}\, t^2\right) & \mbox{ for } m \gg 1/\delta \cr
    \sqrt{N} \frac{t}{\delta} \left(1 - \frac{1}{24} \frac{t^2}{\delta^2}\right) & \mbox{ for } m \ll 1/\delta \hspace{.1in} (\mbox{i.e. } \delta \rightarrow 0)\,.
\end{cases}
\end{align}
We see that in the continuum limit $\delta \rightarrow 0$, the linear growth of complexity diverges proportionally to the inverse of the lattice spacing $\delta^{-1}$. This is in agreement with other calculations of the early-time growth of complexity of a free scalar field using other techniques \cite{Haque:2021hyw}.
In addition, the factor of $\sqrt{N}$ in the complexity (\ref{eq:complexityChain1}), arising from the sum over the $N$-copies of coupled oscillators, indicates that the complexity is proportional to the square root of the volume $\sqrt{\mbox{vol}}$, also in agreement with other techniques.

As with the coupled oscillators, it is of course also possible to diagonalize the Hamiltonian (\ref{eq:Hlattice2}) into the normal modes. In this case, the oscillators all decouple from each other, and the resulting complexity grows strictly linearly with time. However, as discussed in Section \ref{sec:coupled}, the choice of basis gates may be specified by the physical situation of interest, and the normal-mode basis may not be the choice that minimizes the complexity.

We can extend the above approach to consider an \emph{interacting} scalar field theory, such as
\be
\label{eq:interactingScalarH}
\hat H = \int d^{d-1}x \left[\frac{1}{2}\hat \pi^2(x) + \frac{1}{2}\left(\vec{\nabla}\hat \phi\right)^2 + \frac{1}{2} m^2 \hat \phi^2 + \frac{1}{12} \hat\lambda \phi^4 \right]\, .
\ee
The complexity of the interacting scalar field theory has been considered before \cite{Bhattacharyya:2018bbv}, but analysis has been limited by the necessity of using very specific non-Gaussian reference states, and it is not clear that there is a clean decoupling limit $\hat \lambda \rightarrow 0$ that recovers the free scalar field behavior.
Our approach, based on the universal result (\ref{C22}), will not require any specific reference state, and a clear decoupling limit will exist, allowing us to reliably compute the circuit complexity at early times.

As with (\ref{eq:Hlattice}), we put this scalar field on a lattice with spacing $\delta$ and consider it as a set of coupled harmonic oscillators with self-interactions
\be
\label{eq:Hphi4theory}
\hat H_{\rm int} = \sum_{\vec{n}} \left[\frac{\hat P(\vec{n})^2}{2M} + \frac{1}{2} M k^2 \hat X(\vec{n})^2 + \frac{\beta}{2} \sum_i \left(\hat X(\vec{n}) - \hat X(\vec{n}-\hat x_i)\right)^2 + \lambda \hat X(\vec{n})^4 \right]\, .
\ee
with the same definitions for $M,k,\beta$ as in (\ref{eq:Hlattice}), and now we have an anharmonic parameter $\lambda = \hat\lambda/(12\delta^{d+1})$.
The resulting Hamiltonian is a combination of coupled oscillators with fundamental frequency $\alpha^2 = m^2 + \beta/M$, as in Section \ref{sec:coupled}, combined with anharmonic self-interactions similar to (\ref{eq:AnHarmonicH}).
For this system, the ``target gates'' consist of both free and coupled oscillator gates (\ref{eq:coupledHgates}) as well as the self-interaction gates (\ref{eq:AnHarmonicTargetGates}) appearing in the Hamiltonian for each lattice site.
The ``generated gates'' consist of (\ref{AnHarmonicGeneratedGates}) and their coupled oscillator counterparts.

Without performing the detailed analysis of the gates and algebra, however, we know that the general result (\ref{C22}) guarantees that the early-time growth of complexity will be linear, and the behavior will scale as
\be
{\mathcal C}_{\rm interacting} \approx |\omega| t \sim \sqrt{N} \sqrt{a \alpha^2 + b \tilde \beta^2 + c \tilde \lambda^2}\ t\, ,
\label{eq:ComplexityPhi4theory}
\ee
where $\tilde \beta = \beta/(M\alpha)$ (see below (\ref{eq:Hlattice2})) and $\tilde \lambda = \lambda/(M^2 \alpha^2)$ (see below (\ref{eq:AnHarmonicH}),
and $a,b,c$ are some numerical coefficients arising from rewriting (\ref{eq:Hphi4theory}) in terms of the target gates, whose exact values will not be relevant here.
The dominant behavior of the linear growth of complexity is then controlled by either the free oscillator $\alpha$, the coupled oscillator interaction $\tilde\beta$, or the anharmonic self-interaction $\tilde \lambda$.
Note that the limit of $\tilde \lambda \rightarrow 0$ for fixed lattice spacing will lead to a smooth limit of the free scalar field, which is something that is difficult and more subtle using other techniques for studying the circuit complexity of the quartic self-interaction \cite{Bhattacharyya:2018bbv}.
Specifically, the strength of the ``coupled oscillators'' arising in the free scalar field will be controlled by the dimensionless quantity (see (\ref{eq:coupledStrength}))
\be
\frac{\tilde \beta}{\alpha} = \frac{\beta}{M (k^2 + 1/\delta^2)} = \frac{1}{1+m^2\delta^2}\, ,
\ee
which approaches ${\mathcal O}(1)$ in the continuum limit $\delta \rightarrow 0$.
On the other hand, the strength of the self-interactions will be controlled by the dimensionless quantity
\be
\label{eq:Phi4theorySelfInteractionStrength}
\frac{\tilde \lambda}{\alpha} = \frac{\lambda}{M^2\alpha^3} = \frac{\hat\lambda}{12 \delta^{d-1}(m^2 + 1/\delta^2)^{3/2}} \approx \begin{cases}
    \frac{\hat \lambda}{12 \delta^{d-1} m^3} & \mbox{ for } m \gg 1/\delta \cr
    \frac{\hat \lambda}{12 \delta^{d-4}} & \mbox{ for } m \ll 1/\delta
\end{cases}
\ee
in terms of the original coupling constant $\hat \lambda$ of the interacting field theory.
Interestingly, for $d < 4$ the effective strength of the anharmonic term vanishes in the continuum limit $\delta \rightarrow 0$, so the complexity of the system is dominated by the free scalar field result (\ref{eq:complexityChain1}) ${\mathcal C}\sim \sqrt{N}\, t/\delta$
in this limit.
For $4 \leq d < 6$, the self-interaction term in (\ref{eq:ComplexityPhi4theory}) will diverges in the continuum limit, but not as fast as the free scalar field contribution, so again the complexity is dominated by the free scalar field result.
Only for $d \geq 6$ does the contribution of the self-interaction dominate the complexity (\ref{eq:ComplexityPhi4theory}) in the continuum limit, with ${\mathcal C}\sim \sqrt{\hat \lambda}\, t/\delta^{(d-4)/2}$.
Interestingly, this implies that for all $d < 6$ scalar field theories, both free and interacting, the early-time growth in complexity is controlled entirely by divergences from the continuum limit of the \emph{free theory}.
{ It may be surprising that the early-time growth of complexity of scalar fields is insensitive to self-interactions for $d<6$; after all, complexity is expected to be a probe of strongly-interacting physics. Instead, we see a more universal behavior of the complexity of scalar field theories at early times, which is dominated by divergences of the free theory in the continuum limit. We do expect, however, that strong self-interactions should be important at late times, but this regime is beyond the control of our approximation.}

In addition to the linear growth of complexity, following (\ref{C22}) there will also be a subleading term due to the non-trivial algebra gates of the interacting lattice of oscillators subtracting from this linear growth. As with the linear growth, the contributions of the free and interacting parts of the theory will be controlled by the same dimensionless parameters as discussed above. Though the details of the calculation are quite involved and challenging, we can already anticipate some of the general results from our earlier results on the coupled and anharmonic oscillator and (\ref{C22}).
In the limit of weak self-interaction $\tilde \lambda$, we expect both the linear growth rate of complexity and the time scale at which corrections to the linear growth appear to be controlled by the lattice spacing $\delta^{-1}$.
For stronger self-interaction $\tilde \lambda$, we still expect the lattice spacing to control both of these scales for $d < 6$. For $d \geq 6$, however, the inverse of the self-interaction parameter $\tilde\lambda^{-1}$ will control both timescales, as in Section \ref{sec:AnharmonicOscillator}.

\section{Discussion}

The quantum circuit complexity of an operator is an interesting quantity in quantum information theory because it can provide information about how difficult it is to construct a unitary transformation from a given set of gates, and may provide insight into behaviors such as quantum chaos.
In this paper, we considered the early-time behavior for continuous quantum circuit complexity of the unitary time-evolution operator $\hat {\mathcal U}_{\rm target} = {\rm exp}\left[-it\hat H\right]$ for an arbitrary time-independent Hamiltonian $\hat H$.
In addition to the Hamiltonian $\hat H$, a calculation of the quantum circuit complexity also requires a choice of Hermitian gates $\{\hat {\mathcal O}_I\}$ in which the Hamiltonian is decomposed as $\hat H = \omega^I \hat {\mathcal O}_I$, and a ``cost'' metric $G_{IJ}$ that characterizes the difficulty in building the target unitary with the gates $\{\hat {\mathcal O}_I\}$.

Without making any assumptions on the form of the Hamiltonian, gates, or cost metric, 
we showed in Section \ref{sec:pert} that the complexity of $\hat {\mathcal U}_{\rm target}$ takes a universal form at early times
\be
{\mathcal C} \approx t |\omega| \left(1-\frac{1}{24} \frac{|\Omega_2|^2}{|\omega|^2}\, t^2\right)\, ,
\label{eq:C22Discussion}
\ee
where $|\omega| = \sqrt{G_{IJ} \omega^I \omega^J}$ is the magnitude of the Hamiltonian decomposed along the gate set.
The quantity $|\Omega_2|$ is the magnitude of a real vector that depends on the components of $\omega^I$ and the structure constants $f_{IJ}^K$ of the algebra of the gates, $[\hat {\mathcal O}_I, \hat {\mathcal O}_J] = i f_{IJ}^K \hat {\mathcal O}_K$.
This universal form for the complexity at early times has several important features.
First, the leading-order behavior of the complexity is universally linear in time confirming other estimates of the early growth of complexity \cite{Susskind_Universality}, with the rate of change of the complexity given simply by the magnitude of the Hamiltonian when decomposed along the gate set (and weighted by the cost metric).
Second, the sign of the sub-leading term in (\ref{eq:C22Discussion}) is fixed and negative, so that the complexity of unitary time evolution at early times is \emph{bounded by linear growth} for any system, including cases in which there are interactions. 
The origin of this subleading, subtractive, term can be traced to the algebra of the gates; for example, when successive gates are applied, their combined effect can be to generate new, additional terms through their commutators, as in
\be
e^{-i \beta^I \hat {\mathcal O}_I} e^{-i \gamma^I \hat{\mathcal O}_I}  = e^{-i (\beta^I + \gamma^I)\hat {\mathcal O}_I + \beta^I\gamma^J [\hat {\mathcal O}_I, \hat {\mathcal O}_J] + ...} = e^{-i (\beta^I + \gamma^I)\hat {\mathcal O}_I - \frac{i}{2}\beta^I\gamma^J f_{IJ}^K \hat {\mathcal O}_K + \ldots} 
\ee
where the $\ldots$ represent terms with higher commutators of the gates. 
Through the algebra and the structure constants $f_{IJ}^K$, a circuit can generate additional terms, reducing its length, leading to the suppression of the linear growth of complexity in (\ref{eq:C22Discussion}) through the $|\Omega_2|$ term.
In special circumstances, however, $|\Omega_2| = 0$, so the linear growth persists for late times as well (though the growth may ultimately be limited by global periodicity constraints).
We showed how, in the presence of one or more ``hard'' directions in which the cost metric $G_{IJ}^{(\rm hard)} \sim \mu \gg 1$ is large, the linear growth of the complexity scales as the square root of the cost fact ${\mathcal C} \sim \sqrt{\mu}\ t |\omega_{\rm hard}|$, leading to an initially rapid growth in complexity, which is however quickly shut off by the algebra effect discussed above.

We explored the application of the universal result (\ref{eq:C22Discussion}) to several examples, including Hamiltonians constructed from gates satisfying algebras ranging from $\mathfrak{su}(2^N)$ to $\mathfrak{su}(1,1)$, realizing the harmonic oscillator, matching the universal result to known results in the literature and generalizing these results to arbitrary cost metrics $G_{IJ}$.
We also demonstrated how our approach can be used to calculate the complexity for coupled oscillators, finding that while the linear growth of coupled oscillators is controlled by the fundamental frequency of the oscillators, suppression of the linear growth is controlled by the coupling between the oscillators, allowing the linear and higher-order terms to be parametrically separable.
An advantage of our universal result is that since it only requires the algebra of gates, we can calculate the complexity for systems that other approaches are not well-suited to handle.
We demonstrated this power by calculating the complexity for the anharmonic oscillator and showing how the strength of the anharmonic term controls the suppression of linear growth at early times.
Using these results, we were able to extend our analysis to study the complexity of free and self-interacting $\hat \lambda \phi^4$ scalar field theories in $d$ spacetime dimensions by discretizing them on a lattice with spacing $\delta$. While the general behavior for the complexity of the scalar field theories follows the universal result (\ref{eq:C22Discussion}), we were able to show how the linear growth of complexity, and the suppression to that linear growth through $|\Omega_2|$, scales with $\delta$ in the continuum limit as well as the self-interaction coupling constant $\hat \lambda$. Surprisingly, we found that for $d \leq 6$ spacetime dimensions, the initial linear growth in complexity is independent of any $\hat \lambda \phi^4$ self-interaction in the continuum limit, scaling as ${\mathcal C} \sim \sqrt{N} t/\delta$ for $N \rightarrow \infty$ lattice sites and $\delta \rightarrow 0$ spacing. It would be interesting to study  whether it is true more generally that the early-time growth of complexity for strongly coupled systems is dominated by the free theory.
%further to make connections with any universal expectations of the growth of complexity for strongly coupled systems being dominated by the free theory at early-times.

Looking forward, it would be interesting to use the universal result found here to investigate models that allow tuning between integrable and chaotic behavior, since it is possible that the transition into quantum chaos may be observable in the behavior of the higher-order in time corrections to the early-time complexity.
This current work does not focus on the holographic perspective. Instead, we explored complexity purely within the quantum mechanical and field theory framework. By using the approach outlined in this paper, however, one could investigate different interacting field theories with large coupling that have a gravity dual description. Progress in understanding complexity for interacting quantum field theories can provide valuable insights into holographic scenarios. 
Finally, we note that the results here may be a useful tool in comparing and contrasting different approaches to complexity to each other.
We hope to explore these and other interesting topics in future work.

\section*{Acknowledgment}
We would like to thank Chandan Jana for discussions. G.~J.~is supported by the “Quantum Technologies for Sustainable Development'' project from the National Institute for Theoretical and Computational Sciences (NITheCS). 

\appendix

\section{Derivation of EA equation}
\label{app:EulerArnoldDerivation}

In this appendix, we review the derivation of geodesics in the space of an arbitrary group manifold $G$. Denoting an arbitrary unitary by $U$, we choose a gate basis on the tangent space (i.e. the Lie algebra) as the set $\{\mathcal{O}_I\}$. A path is given by $U(s)$ and we parameterize the path by $s\in [0,1] $. In order to define the distance we need a metric on the group manifold.
A simple metric would be 
\begin{equation}
    d\ell^2=\operatorname{tr}(\Omega^\dagger \mathbb{J}\Omega)\, ds^2\,.
\end{equation}
The matrix quantity $\Omega$  is the velocity on the group manifold and is related to group elements along the path by $\Omega=i\dot{U}U^{-1}$ and $\mathbb{J}$ is some fixed matrix in the definition of the inner product.  
A dot indicates a derivative with respect to the ``circuit parameter'' $s$. 
Expanding $\Omega$ in the basis $\mathcal{O}_I$, we have $\Omega=V^I(s)\mathcal{O}_I$. 
The corresponding metric elements in this basis will then be given by $\operatorname{tr}(\mathcal{O}^{\dagger}_I \mathbb{J}\, \mathcal{O}_J)=G_{IJ}$. 
In the case of an orthonormal basis of gates, and if we choose $\mathbb{J}$ to be the identity matrix then the metric will be $G_{IJ}=\delta_{IJ}$. 
On the other hand if for a given group if we use the adjoint representation for the basis gates $\mathcal{O}_I$, then the metric will be the Cartan-Killing form \emph{i.e.} $G_{IJ}=K_{IJ}=f^L_{IM}f^{M}_{LJ}$. 

The distance from $U(0)$ to $U(1)$ is given by 
\begin{equation}\label{ac1}
    \ell=\int_{0}^{1}\sqrt{V^IV^JG_{IJ}}\, ds\,.
\end{equation}
To find the minimal geodesic distance we should vary this action and solve the resulting equations of motion. 
But before varying this action let us notice some points. First, note that the variations of the $V^I$ are \emph{induced} from variations of $g$ that are fixed at the endpoints \emph{i.e.} $\delta U(0)=\delta U(1)=0$. Here we fix $U(0)=\mathbb 1$ and $U(1)=\mathcal{U}_{\rm target}$, some fixed target operator. 
Second, for the sake of the equations of motion, we could use the equivalent action 
\begin{equation}\label{ac2}
    S=\tfrac{1}{2}\int_{0}^{1}V^IV^JG_{IJ}\, ds\, .
\end{equation}
Varying this action we simply get
\begin{equation}\label{dac1}
\delta S=\int_{0}^{1}\delta V^IV^JG_{IJ}\, ds\,.
\end{equation}
In order to obtain the equations of motion we should find the relation of $\delta V$ and $\delta \hat U$. From the definition we have:
\begin{equation}
    \label{vdef}
    V^{I}=i G^{IJ}\operatorname{tr}(\mathcal{O}^{\dagger}_{J}\dot{U}U^{-1})\, ,
\end{equation}
where $G^{IJ}$ is the inverse of $G_{IJ}$. 
Variation of both sides leads to
\begin{equation}
    \label{vvar}
    \delta V^{I}=iG^{IJ}\operatorname{tr}(\mathcal{O}^{\dagger}_{J}\delta\dot{U}U^{-1}-\mathcal{O}^{\dagger}_{J}\dot{U}U^{-1} \delta U U^{-1}) \,.
\end{equation}
Now we define  the variable $\eta=i \delta U U^{-1}$ that, like $\delta U$, vanishes at both ends. 
Expanding it in the basis as $\eta=\eta^I \mathcal{O}_I$ and taking a derivative with respect to the circuit parameter $s$ we find
\begin{equation}
    \label{doteta}     
    \dot{\eta}^{I}=i G^{IJ}\operatorname{tr}(\mathcal{O}_{J}^{\dagger}\delta\dot{U}U^{-1}-\mathcal{O}_{J}^{\dagger}\delta{U}U^{-1}  \dot{U} U^{-1}) \,.
\end{equation}
Comparing these equations we find
\begin{equation}\label{ddrule}
    \delta V^{I}=\dot{\eta}^{I}+ f^{I}_{JK}V^K \eta^J\, ,
\end{equation}
where we have used the algebra of the gates $$[\mathcal{O}_J,\mathcal{O}_K]=i f^I_{JK}\mathcal{O}_I.$$ 
Using the \eqref{ddrule} in \eqref{dac1} we find
\begin{equation}\label{dac2}
    \delta S=2\int_{0}^{1}V_I (\dot{\eta}^{I}+ f^{I}_{JK}V^K \eta^J)\, .
\end{equation}
By integration by parts and with the vanishing of $\eta^I$ at the endpoints, we find
\begin{equation}
    \label{eom}
    G_{IJ}\dot{V}^{I}= G_{LK}f^{L}_{IJ}V^{K}V^{J}\, ,
\end{equation}
which is the Euler-Arnold equation for the $V^I$.

\section{Perturbative Solutions for Time-Dependent Hamiltonians}
\label{sec:TimeDepPert}

In the main text, we considered the complexity of the time-evolution operator when the Hamiltonian is time-independent. A generalization of the above procedure to the case of the time-dependent Hamiltonian, however, is straightforward. 
In this case, the time-evolution operator would be 
\begin{equation}
 \hat{\mathcal{U}}(t_{\rm target})=e^{-i \int^t_0 \omega(\tau)^I\hat{\mathcal{O}}_I d \tau}\, .
\end{equation}
The perturbative-in-time expansion of this operator would be
\begin{align}
   \hat{\mathcal{U}}(t)_{\rm target}=&\hat{\mathbf{1}}-i t \omega^I \hat{\mathcal{O}}_I-\frac{t^2}2 (\omega^I \hat{\mathcal{O}}_I)^2-\frac{it^2}2 \omega'^I \hat{\mathcal{O}}_I\nonumber\\
   &+i \frac{t^3}6 (\omega^I \hat{\mathcal{O}}_I)^3-i \frac{t^3}6\omega''^I \hat{\mathcal{O}}_I-\frac{t^3}2 \omega^I \omega'^J \hat{\mathcal{O}}_I\hat{\mathcal{O}}_J-i\frac{t^3}4 \omega^I \omega'^Jf^K_{IJ}\hat{\mathcal{O}}_K\nonumber\\&+O(t^4)\, ,
\end{align}
where by $\omega$, $\omega'$ and $\omega''$, we mean $\omega(0)$, $\omega'(0)$ and $\omega''(0)$ respectively. 
Now we can follow a similar procedure as Section \ref{sec:pert}, although the details are more cumbersome. 
The difference with the time-independent case appears evaluating the solution at the boundary conditions. 
Carefully evaluating the boundary conditions order-by-order, the complexity to ${\mathcal O}(t^3)$ is
	\begin{equation}
		\label{C22int}
		\mathcal{C}\approx  t |\omega|\left(1+\tfrac12 t\frac{\boldsymbol{\omega'}\cdot\boldsymbol{\omega}}{|\omega|^2}+\frac{t^2}{|\omega|^2}\left[- \tfrac1{24}{|\Omega_2|^2}-\tfrac18 \frac{|\boldsymbol{\omega'}\cdot\boldsymbol{\omega}|^2}{|\omega|^2}+\tfrac18 {|\omega'|^2}-\tfrac12 {\boldsymbol{\Omega}_2\cdot\boldsymbol{\omega'}}+\tfrac16 {\boldsymbol{\omega''}\cdot\boldsymbol{\omega}}\right]\right)\, .
	\end{equation}
Clearly, this reduces to the time-independent result (\ref{C22}) when $\omega^I$ is time-independent.
%The important difference is that we also have a $t^2$ correction absent in the time-independent case.

\section{Target and Generated Gates for Anharmonic Oscillator}
\label{app:Anharmonic}

In Section \ref{sec:AnharmonicOscillator}, we introduced a technique for calculating the next-to-leading order in time result for complexity for an anharmonic oscillator of the form
\be
\hat H_{\rm anharm} &=& \frac{1}{2m} \hat p^2 + \frac{m\alpha^2}{2} \hat x^2 + \lambda \hat x^4 \nonumber \\
&=& \left(\alpha + 3 \tilde \lambda \right) \hat e_0 + \frac{1}{2}\tilde \lambda \left(3 \hat e_1 + 3 \hat e_2 + 2 \hat e_3 + \hat e_4\right)\,,
\label{app:AnHarmonicH}
\ee
where $\tilde \lambda \equiv \lambda/(m\alpha^2)$,
and we used the ``target'' gates $\{\hat e_i\}$ (denoted by latin indices)
\begin{eqnarray}
\label{app:AnHarmonicTargetGates}
    \hat e_0 &=& \hat a^\dagger \hat a \,; \\
    \hat e_1 &=&\left(\hat a^{\dagger 2}+\hat a^2\right);\nonumber \\
    \hat e_2 &=& \hat a^{\dagger 2} \hat a^2\,; \nonumber \\
    \hat e_3 &=&\left(\hat a^{\dagger 3} \hat a+\hat a^{\dagger} \hat a^3\right);\nonumber \\
    \hat e_4 &=& \left(\hat a^{\dagger 4}+\hat a^4\right)\, ,\nonumber
\end{eqnarray}
in terms of the (free) raising and lowering operators
\be
\hat a&=&\sqrt{\frac{m\alpha}{2}}\left({\hat {x}}+{\frac{i}{m\omega }}{\hat {p}}\right)\, , \hspace{.4in} \hat a^{\dagger }={\sqrt{\frac{m\alpha}{2}}}\left({\hat {x}}-{\frac{i}{m\omega}}{\hat {p}}\right)\,.
\ee
In addition to the ``target'' gates above, we also need to introduce a natural set of ``generated'' gates $\{\hat e_\alpha\}$ (denoted by greek indices)\footnote{Though not all of these gates will be generated by the algebra of the target gates, they are nonetheless useful for constructing the algebra between the target and generated gates.}
\begin{align}
\label{app:GeneratedGates}
    \hat e_5 & = i\left(\hat a^{\dagger 2} - \hat a^2\right),        & \hat e_8 &= \hat a^{\dagger 6}+\hat a^6\,, \\
    \hat e_6 & = i\left(\hat a^{\dagger 3}\hat a - \hat a^{\dagger} \hat a^3\right),   &      \hat e_9 &= i\left(\hat a^{\dagger 6}-\hat a^6\right), \nonumber \\
    \hat e_7 &= i\left(\hat a^{\dagger 4} - \hat a^4\right),      &    \hat e_{10} &= \hat a^{\dagger 5}\hat a + \hat a^{\dagger} \hat a^5\,, \nonumber \\
    &   &   \hat e_{11} &= i\left(\hat a^{\dagger 5}\hat a - \hat a^{\dagger} \hat a^5\right),\nonumber  \\
    &   &   \hat e_{12} &= \hat a^{\dagger 4}\hat a^2 + \hat a^{\dagger 2}\hat a^4 \,,\nonumber \\
    &   &   \hat e_{13} &= i\left(\hat a^{\dagger 4}\hat a^2 - \hat a^{\dagger 2}\hat a^4\right), \nonumber \\
    &   &   \hat e_{14} &= \hat a^{\dagger 3}\hat a^3\,.\nonumber 
\end{align}
These generated gates include all Hermitian combinations of powers of $\hat a^n, \hat a^{\dagger m}$  for $n+m$ even up to power $n+m = 6$ (combinations of $n+m$ odd will not be generated).
The algebra between the target gates $\{\hat e_i\}$ and the generated gates $\{\hat e_\alpha\}$ will determine the non-zero contributions to the complexity at next-to-leading order in time.

The algebra of the target and generated gates determines the structure constants through
$\left[\hat e_I,\hat e_J\right] = i f_{IJ}^K\hat e_K$.
The algebra constructed by commuting the target gates with each other is
\begin{align}
\left[\hat e_0,\hat e_1\right] &= - 2 i \hat e_5\,, & \left[\hat e_1,\hat e_2\right] &= - 2 i \hat e_5 - 4 i \hat e_6\,, \\
\left[\hat e_0,\hat e_2\right] &= 0\,, &  \left[\hat e_1,\hat e_3\right] &= - 2 i \hat e_7\,, \nonumber \\
\left[\hat e_0,\hat e_3\right] &= -2 i \hat e_6\,, &  \left[\hat e_1,\hat e_4\right] &= - 12 i \hat e_5 - 8 i \hat e_6\,, \nonumber \\
\left[\hat e_0,\hat e_4\right] &= -4 i \hat e_7\,, &  \left[\hat e_2,\hat e_3\right] &= - 6 i \hat e_6 - 4 i \hat e_{13}\,, \nonumber \\
\left[\hat e_3,\hat e_4\right] &= -24 i \hat e_5 - 36 i \hat e_6 - 4 i \hat e_9 - 12 i \hat e_{13}\,, &  \left[\hat e_2,\hat e_4\right] &= - 12 i \hat e_7 - 8 i \hat e_{11}\,. \nonumber
\end{align}
Interestingly, the structure constants among the target gates vanish, $f_{ij}^k = 0$, but we have non-zero structure constants of the form $f_{ij}^\alpha$, in which two target gates commute to a generated gate. Because of the structure of the vector $\Omega_2^I$, however, this algebra will not contribute to the complexity at next-to-leading order, and only serves to motivate the introduction of the generated gates.

Instead, contributions to the complexity arise from structure constants $f_{i\alpha}^k$ that involve commutators between a target gate and a generated gate leading to a target gate, of the form
\be
\left[\hat e_i, \hat e_\alpha\right] = i f_{i\alpha}^k \hat e_k + \ldots
\label{app:CommutatorContribution}
\ee
where the $\ldots$ could be additional terms involving generated gates.
For commutators involving $\hat e_0$, we can use the relation
\be
\left[a^{\dagger} a, a^{\dagger k} a^{\ell} \pm a^{\dagger \ell} a^k\right]=(k-\ell)\left(a^{\dagger k} a^{\ell} \mp a^{\dagger \ell} a^k\right)\,,
\ee
to find the algebra
\begin{align}
\label{app:e0Algebra}
    \left[\hat e_0,\hat e_5\right] & = 2 i \hat e_1\,, & \left[\hat e_0,\hat e_6\right] & = 2 i \hat e_3 \,,
     & \left[\hat e_0,\hat e_7\right] & = 4 i \hat e_4\,,
\end{align}
with all other commutators either not being relevant (i.e.~the commutator does not contribute a term of the form (\ref{app:CommutatorContribution})) or zero.
For the other target gates, we simply need to calculate the relevant commutators
\begin{align}
\label{app:targetGeneratedAlgebra}
    \left[\hat e_1, \hat e_5 \right] & = 4 i + 8 i \hat e_0 & \left[\hat e_2, \hat e_5 \right] & = 2 i \hat e_1 + 4 i \hat e_3\\
    \left[\hat e_1, \hat e_6 \right] & = 12 i \hat e_0 + 12 i \hat e_2 - 2 i \hat e_4 & \left[\hat e_2, \hat e_6 \right] & = 6 i \hat e_3 + 4 i \hat e_{12} \nonumber \\
    \left[\hat e_1, \hat e_7 \right] & = 12 i \hat e_1 + 8 i \hat e_3 & \left[\hat e_2, \hat e_7 \right] & = 12 i \hat e_4 + 8 i \hat e_{10 \nonumber }\\ 
    \left[\hat e_1, \hat e_9 \right] & = 30 i \hat e_4 + 12 i \hat e_{10} & \left[\hat e_3, \hat e_5 \right] & = 12 i \hat e_0 + 12 i \hat e_2 + 2 i \hat e_4 \nonumber \\
    \left[\hat e_1, \hat e_{11} \right] & = 20 i \hat e_3 - 8 i \hat e_8 + 10 i \hat e_{12} & \left[\hat e_3, \hat e_6 \right] & = 12 i \hat e_0 + 36 i \hat e_2 + 16 i \hat e_{14} \nonumber \\
    \left[\hat e_1, \hat e_{13} \right] & = 24 i \hat e_2 -2 i \hat e_4 - 4 i \hat e_{10} + 16 i e_{14} & \left[\hat e_3, \hat e_7 \right] & = 24 i \hat e_1 + 36 i \hat e_3 + 4 i \hat e_8 + 12 i \hat e_{12}  \nonumber \\
    \left[\hat e_4, \hat e_5 \right] & = 12 i \hat e_1 + 8 i \hat e_4 & \left[\hat e_3, \hat e_9 \right] & = 120 i \hat e_4 + 90 i \hat e_{10} + \mbox{higher-order}  \nonumber \\
    \left[\hat e_4, \hat e_6 \right] & = 24 i \hat e_1 + 36 i \hat e_3 - 4 i \hat e_8 + 12 i \hat e_{12} & \left[\hat e_3, \hat e_{11} \right] & = 60 i \hat e_3 + 60 i \hat e_{12} + \mbox{higher-order}  \nonumber \\
    \left[\hat e_4, \hat e_7 \right] & = 48 i + 192 i \hat e_0 + 144 i \hat e_2 + 32 i \hat e_{14} & \left[\hat e_3, \hat e_{13} \right] & = 48 i \hat e_2 - 6 i \hat e_{10} + 72 i \hat e_{14} \nonumber \\
    & & & + \mbox{higher-order}  \nonumber \\
    \left[\hat e_4, \hat e_9 \right] & = 360 i \hat e_1 + 480 i \hat e_3 + 180 i \hat e_{12} & \left[\hat e_4, \hat e_{11} \right] & = 240 i \hat e_0 + 480 i \hat e_2 + 240 i \hat e_{14} \nonumber \\
    & + \mbox{higher-order}  & & + \mbox{higher-order}  \nonumber \\
    \left[\hat e_4, \hat e_{13} \right] & = 24 i \hat e_1 + 96 i \hat e_3 + 12 i \hat e_8 + 72 i \hat e_{12} & & \nonumber \\
    &+ \mbox{higher-order} & &\nonumber 
\end{align}
where ``higher-order'' denotes a (Hermitian) contribution of the form $\hat a^{\dagger n}\hat a^m$ where $m+n = 8$, and thus is not relevant for this analysis.

From this algebra, we can extract the structure constants and calculate the quadratic vector
\be
\Omega_2^I = G^{IL} f_{LK}^J \omega^K G_{JM} \omega^M\, .
\label{app:AnHarmonicOmega2}
\ee
As discussed in Section \ref{sec:AnharmonicOscillator}, assuming a diagonal, equal-cost metric $G_{IJ} = \delta_{IJ}$, the structure of (\ref{app:AnHarmonicOmega2}) simplifies so that the components along the target gate directions take the form
\be
\Omega_2^i = G^{i\ell} f_{\ell K}^J \omega^K G_{JM} \omega^M = G^{i\ell} f_{\ell k}^j \omega^k G_{jm} \omega^m\, .
\ee
Since we do not have any non-zero structure constants of this form, these components vanish identically $\Omega_2^i = 0$.
The components of (\ref{app:AnHarmonicOmega2}) along the \emph{generated} gates directions takes the form
\be
\Omega_2^\alpha = G^{\alpha \beta} f_{\beta K}^J \omega^K G_{JM} \omega^M = G^{\alpha\beta} f_{\beta k}^j \omega^k G_{jm} \omega^m\, .
\label{app:AnHarmonicOmega2Generated}
\ee
For the algebra (\ref{app:e0Algebra}) and (\ref{app:targetGeneratedAlgebra}), it is straightforward to find the $f_{\beta k}^j = -f_{k\beta}^j$ and calculate (\ref{app:AnHarmonicOmega2Generated}), which has components
\begin{align}
\Omega_2^5 & = -27\alpha \tilde \lambda - \frac{243}{2} \tilde \lambda^2\,, & \Omega_2^6 & = 4 \alpha \tilde \lambda - \frac{83}{2} \tilde \lambda^2 \,,\\
\Omega_2^7 & = -98 \alpha \tilde \lambda - 444 \tilde \lambda^2\,, & \Omega_2^8 &= 0 \,,\nonumber \\
\Omega_2^9 & = -\frac{1185}{2} \tilde \lambda^2\,, & \Omega_2^{10} & = 0\,, \nonumber \\
\Omega_2^{11} & = -120 \alpha \tilde \lambda - 810 \tilde \lambda^2\,, & \Omega_2^{12} & = 0\,, \nonumber \\
\Omega_2^{13} & = -\frac{381}{2} \tilde \lambda^2\,, & \Omega_2^{14} & = 0 \,.\nonumber
\end{align}
From these components, it is straightforward to calculate $|\Omega_2|^2 = G_{IJ} \Omega_2^I \Omega_2^J$, which appears in the subleading-in-time expression for complexity
\be
\label{app:Omega2Result}
|\Omega_2|^2 = \tilde \lambda^2 \left(a\,\alpha^2 + b\, \alpha\, \tilde \lambda + c\, \tilde\lambda^2\right)\, ,
\ee
where $a,b,c$ are pure numbers with values $a = 24\, 749\, , b = 287\, 653\, , c = 1\, 257\, 067$.

It is also easy to extend these calculations for a cost-metric that is diagonal, but gives a higher cost $\mu$ to the ``generated'' gates compared to the target gates, e.g.
\be
G_{IJ} = \begin{pNiceMatrix}
1 &  & & & & \\
    & \Ddots & & & & \\
    & & 1 & & & \\
    & & & \hspace{.1in}\mu & & \\
    & & & & \Ddots & \\
    & & & & & \mu
\end{pNiceMatrix}\,.
\ee
Since the vector $\Omega_2^I$ only has components along the ``generated'' gates direction, this simply multiplies the result of (\ref{app:Omega2Result}) by an inverse factor of the weight $\mu$:
\be
|\Omega_2|^2 = \frac{\tilde \lambda^2}{\mu} \left(a\,\alpha^2 + b\, \alpha\, \tilde \lambda + c\, \tilde\lambda^2\right)\,,
\ee
where $a,b,c$ have the same numerical values as before.

{
An alternative cost metric is to assign a ``cost'' for each gate associated with powers of the raising and lowering operators.
In particular, gates $\hat e_0,\hat e_1$, which are second order in $\hat a,\hat a^\dagger$, will have a cost factor of unity; gates $\hat e_2$ through $\hat e_7$, which are fourth order in $\hat a, \hat a^\dagger$, will have a cost factor $\mu$, even though some of them are target gates and some of them are generated gates; and finally, gates $\hat e_8$ through $\hat e_{14}$ will have a cost factor $\mu^2$, since they are sixth order in $\hat a,\hat a^\dagger$.
In this case, we obtain at leading order in $\mu$,
\be
|\Omega_2|^2 \approx a \mu \tilde\lambda^4
\ee
where $a = 149629/4$ is some numerical constant.
}

\bibliographystyle{utcaps}

\bibliography{reference}

\end{document}